\documentclass[10pt,journal,a4paper,twoside,twocolumn]{IEEEtran}
\usepackage{cite}
\usepackage{float}
\usepackage{stfloats}
\usepackage{fancyhdr}
\usepackage{color}
\usepackage[inline]{enumitem}
\usepackage{amsmath,amsfonts,amssymb}
\usepackage{graphicx}
\usepackage{algorithm}
\usepackage{algorithmic}
\usepackage{epsfig}
\usepackage{subfigure}
\usepackage{epstopdf}
\usepackage{txfonts}
\usepackage{bm}
\usepackage{array}

\usepackage{bbm}

\newtheorem{Lemma}{Lemma}
\newtheorem{Rem}{Remark}

\begin{document}
	\title{Reconfigurable Intelligent Sensing Surface aided Wireless Powered Communication Networks: A Sensing-Then-Reflecting Approach
	}
	
	\author{\IEEEauthorblockN{
			Cheng Luo, \emph{Student Member, IEEE}, Jie Hu, \emph{Senior Member, IEEE}, Luping Xiang, \emph{Member, IEEE} and Kun Yang, \emph{Fellow, IEEE}
            }
			\\
        \thanks{Cheng Luo, Jie Hu and Luping Xiang are with the School of Information and Communication Engineering, University of Electronic Science and Technology of China, Chengdu, 611731, China, email: chengluo@std.uestc.edu.cn; hujie@uestc.edu.cn; luping.xiang@uestc.edu.cn.}
        \thanks{Kun Yang is with the School of Computer Science and Electronic Engineering, University of Essex, Colchester CO4 3SQ, U.K., email: kunyang@essex.ac.uk.}
	}

	\maketitle
	
	\thispagestyle{fancy} 
	\lhead{} 
	\chead{} 
	\rhead{} 
	\lfoot{} 
	\cfoot{} 
	\rfoot{\thepage} 
	\renewcommand{\headrulewidth}{0pt} 
	\renewcommand{\footrulewidth}{0pt} 
	\pagestyle{fancy}

    \rfoot{\thepage} 

\begin{abstract}
    This paper presents a reconfigurable intelligent sensing surface (RISS) that combines passive and active elements to achieve simultaneous reflection and direction of arrival (DOA) estimation tasks. By utilizing DOA information from the RISS instead of conventional channel estimation, the pilot overhead is reduced and the RISS becomes independent of the hybrid access point (HAP), enabling efficient operation. Specifically, the RISS autonomously estimates the DOA of uplink signals from single-antenna users and reflects them using the HAP's slowly varying DOA information. During downlink transmission, it updates the HAP's DOA information and designs the reflection phase of energy signals based on the latest user DOA information. The paper includes a comprehensive performance analysis, covering system design, protocol details, receiving performance, and RISS deployment suggestions. We derive a closed-form expression to analyze system performance under DOA errors, and calculate the statistical distribution of user received energy using the moment-matching technique. We provide a recommended transmit power to meet a specified outage probability and energy threshold. Numerical results demonstrate that the proposed system outperforms the conventional counterpart by 2.3 dB and 4.7 dB for Rician factors $\kappa_h=\kappa_G=1$ and $\kappa_h=\kappa_G=10$, respectively.
\end{abstract}

\begin{IEEEkeywords}
    Reconfigurable intelligent sensing surface, wireless-powered communication network, direction of angle estimation, performance analysis.
\end{IEEEkeywords}

\section{Introduction}
\IEEEPARstart{R}{econfigurable} intelligent surface (RIS), a promising innovation in the domain of wireless information transfer (WIT) and wireless energy transfer (WET), offers the potential to augment energy efficiencies, coverage, and security \cite{liu2021reconfigurable}. A RIS is typically characterized by a planar surface equipped with a multitude of cost-effective, passive reflecting elements. These components can independently manipulate the phase of incoming electromagnetic signals to satisfy specific functional and performance needs, thus reducing the impact of unfavorable wireless propagation environments and introducing an extra line-of-sight (LoS) channel.

Owing to these benefits, RIS-assisted WIT and WET systems have been the subject of considerable academic attention. Several research initiatives have explored the joint design of active beamforming and passive RIS reflecting phases, as documented in \cite{jointlydesign1,jointlydesign2,jointlydesign3}. Concurrently, other investigations, such as \cite{discretephase1,discretephase2}, have scrutinized the more pragmatic scenario where the RIS has a finite number of bits for phase adjustment. A novel model incorporating coupled reflecting phase and amplitude in RIS was proposed in \cite{coupledPhaseAmp}, elucidating further optimization of the RIS. In \cite{SecureWIPT}, a low-complexity alternating optimization algorithm was introduced, demonstrating energy harvesting nearly twice as efficient as systems without RIS while maintaining secrecy rate requirements. Moreover, \cite{singlesensorDOA} proposed a single sensor direction of arrival (DOA) estimation scheme leveraging compressed sensing, indicating promise for future applications due to lower hardware complexity and cost. Furthermore, the integration of RIS with other emerging technologies has been explored, including spatial electromagnetic wave frequency mixing \cite{frequencyadjustment2}, active RIS \cite{activeIRS2}, robust beamforming \cite{robustBF1}, joint communication and sensing design \cite{JointCommSensing}, in addition to simultaneous wireless information and power transfer (SWIPT) \cite{swipt1,swipt2} and wireless-powered communication network (WPCN) \cite{wpcn,wpcn2}.

The aforementioned research underscores the prospective advantages of incorporating RIS into future WIT/WET applications. However, the deployment of RIS in WET encounters a formidable obstacle in obtaining channel state information (CSI) between the RIS and its affiliated PBs/users \cite{wqqtutorial}. The lack of signal processing functionalities within the RIS compounds the complexity in estimating channels from the HAP to the RIS and eventually to the users. Further complicating the matter is the pilot overhead, which increases in direct proportion to the number of RIS elements and users, as corroborated by \cite{pilotoverhead1,pilotoverhead2}, thereby making channel estimation unfeasible. Moreover, the absence of active radio frequency (RF) links in economical reflecting elements precludes pilot transmission for channel estimation.

In response to the aforementioned predicaments, research endeavors in recent years have sought to address the RIS channel estimation problem. A notable study by \cite{csi_deeplearning} offers a solution to the RIS channel estimation dilemma by integrating deep learning and compressed sensing methodologies with randomly distributed active sensors, thereby facilitating channel estimation with negligible pilot overhead. Additionally, a novel technique proposed in \cite{sensingirs_dll} suggests equipping a power sensor behind each RIS element, which allows the signal to be superposed in the same phase at the receiver through interference observation. Furthermore, the optimal precoder and RIS phase shift are determined based on the collection of a substantial volume of empirical observations, as elaborated in \cite{blindBF}. A location information-aided RIS system was also proposed in \cite{locinformation}, utilizing location information to design the transmission and RIS beamforming strategies. \cite{StatisticalCSI} investigated joint beamforming and RIS phase shift design using statistical CSI, achieving positive results. Partial related works are presented in Table \ref{table:relatedworks}.
\begin{table*}[]
    \centering
    \caption{Relevant works on communication assisted by RIS and sensing.}
    \begin{tabular}{|m{2.9 cm}<{\centering}|m{2.5 cm}<{\centering}|m{2.8 cm}<{\centering}|m{1.3 cm}<{\centering}|m{3 cm}<{\centering}|}
    \hline
    \textbf{Sensing Method} &  \textbf{HAP control requirements}&  \textbf{Simultaneous Sensing and Comm.}& \textbf{Channel Model} & \textbf{Performance analysis with sensing error} \\
    \hline
    Our Proposed&No&Yes&Rician&Yes\\
    \hline
    CS\cite{singlesensorDOA}&Yes&No&LoS&No\\
    \hline
    GPS-based\cite{locinformation}&Yes (small-scale data transmission)&Yes&LoS&Yes\\
    \hline
    ESPRIT+MUSIC\cite{huxiaol3}&Yes& No& LoS&No\\
    \hline
    2D-DFT+AO\cite{2dDFT}&Yes&No&LoS&No\\
    \hline
    EM nature\cite{sensingirs_dll}&Yes&Yes&Rayleigh&No\\
    \hline
    \end{tabular} \label{table:relatedworks}
\end{table*}

In summary, the extant research on WIT and WET systems, aided by RIS, is subject to several limitations:
\begin{itemize}
\item The necessity for CSI renders the pilot overhead of RIS-assisted systems unfeasible, particularly for large-scale RIS.
\item The design of the reflecting phase is excessively dependent on the feedback link between RIS and HAP, significantly impeding the potential for scale expansion, extensive deployment, and migration of RIS.
\item The predominant focus of current research is on the design of LoS channels, with Rician channels receiving limited analytical attention.
\end{itemize}

In this paper, we detail the design and analysis of a reconfigurable intelligent sensing surface (RISS) assisted WPCN system. Our key contributions can be encapsulated as follows:

\begin{itemize}
\item Initially, we introduce a transmission protocol and system model for WPCN with DOA estimation assistance. We derive a closed-form expression for the expected energy reception with our proposed scheme, while we provide a tight upper bound on the communication ergodic spectral efficiency. Utilizing the derived closed-form results, we examine the necessary conditions for achieving transmitter antenna gains and offer deployment suggestions for the RISS under this scheme. Then we deduce the distribution of user received energy and SNR through the moment-matching method. Based on the derived distribution, we obtain the ergodic spectral efficiency in closed-form. We also obtain the minimum transmit power that satisfies the outage probability constraint, given a certain receive energy threshold. Numerical results indicate that the closed-form solutions acquired using moment-matching are consistent with Monte Carlo simulations, and the outage probability performance of the receiver can be effectively managed by adjusting the transmit power.
\item We analyze and model the errors in DOA estimation. Primarily, based on DOA estimation experiments, the estimation errors in the spatial phase differences $(u,v,z)$ is modeled as a Gaussian distribution. We derive a closed-form solution for the errors in the spatial phase differences $(u,v,z)$, and by linking it to the errors in angle estimation through fitting, we achieve a final closed-form solution in the angle domain. Numerical results underscore that our closed-form solution closely aligns with Monte Carlo simulation results. This concordance is vital for the performance evaluation of practical models.
\item Our proposed scheme allows the RISS to operate autonomously, relying solely on angle information derived from DOA estimation for the independent design of the reflecting phase, thus detaching RISS from the HAP. This aspect carries significant implications for the future deployment and migration of RIS.
\item Our proposed scheme conserves pilot overhead while introducing only negligible energy and computational costs for a modest number of active elements. Numerical results substantiate that our scheme is more practical for large-scale RIS systems and surpasses the traditional full CSI counterpart by 2.3 dB when the Rician factors in both the HAP-to-RISS channel and the RISS-to-user channel are 1. Moreover, the performance gain is as high as 4.7 dB, when the Rican factors in both channels are 10. 
\end{itemize}

The remainder of this paper is organized as follows. Section \ref{sec:sec2} provides an overview of the system model and transmission protocol, while Section \ref{sec:sec3} analysis the performance of proposed scheme and obtain the closed-form expressions. The numerical results are presented in Section \ref{sec:sec4}, and Section \ref{sec:sec5} summarizes the findings and conclusions of this study.

 \emph{Notation:} $\mathbf{I}_{M}$ represents the $M$-dimensional identity matrix, and $\mathbf{1}_{M\times N}$ represents the $M\times N$ matrix with all ones. The notation $[\cdot]_i$ and $[\cdot]_{i,j}$ refers to the $i$-th element of a vector and the $(i,j)$-th element of a matrix, respectively. The imaginary unit is denoted by $\mathbbm{i}=\sqrt{-1}$. The Euclidean norm and absolute value are denoted by $||\cdot||$ and $|\cdot|$, respectively. The function $\text{diag}(\cdot)$ creates a diagonal matrix. The operators $(\cdot)^\mathrm{T}$ and $(\cdot)^\mathrm{H}$ represent the transpose and conjugate transpose, respectively. The operator $\Re(\cdot)$ extracts the real component of a complex number, while $\Im(\cdot)$ extracts the imaginary component. The mathematical expectation and variance are denoted by $\mathbb{E}(\cdot)$ and $\mathbb{D}(\cdot)$, respectively. The notation $Z\sim\mathcal{X}^2_{a, b}$ represents the chi-square distribution with $a$ degrees of freedom and the common variance $b$ of its elements. The mean value and variance of $Z$ are $\mathbb{E}(Z)=ab$ and $\mathbb{D}(Z)=2ab^2$, respectively. Finally, the notation $\mathcal{CN}$ and $\mathcal{N}$ represent the circularly symmetric complex gaussian distribution and gaussian distribution, respectively.

\section{System Model and Transmission Protocols}\label{sec:sec2}
    \begin{figure}
        \centering
        \includegraphics[width=0.8\linewidth]{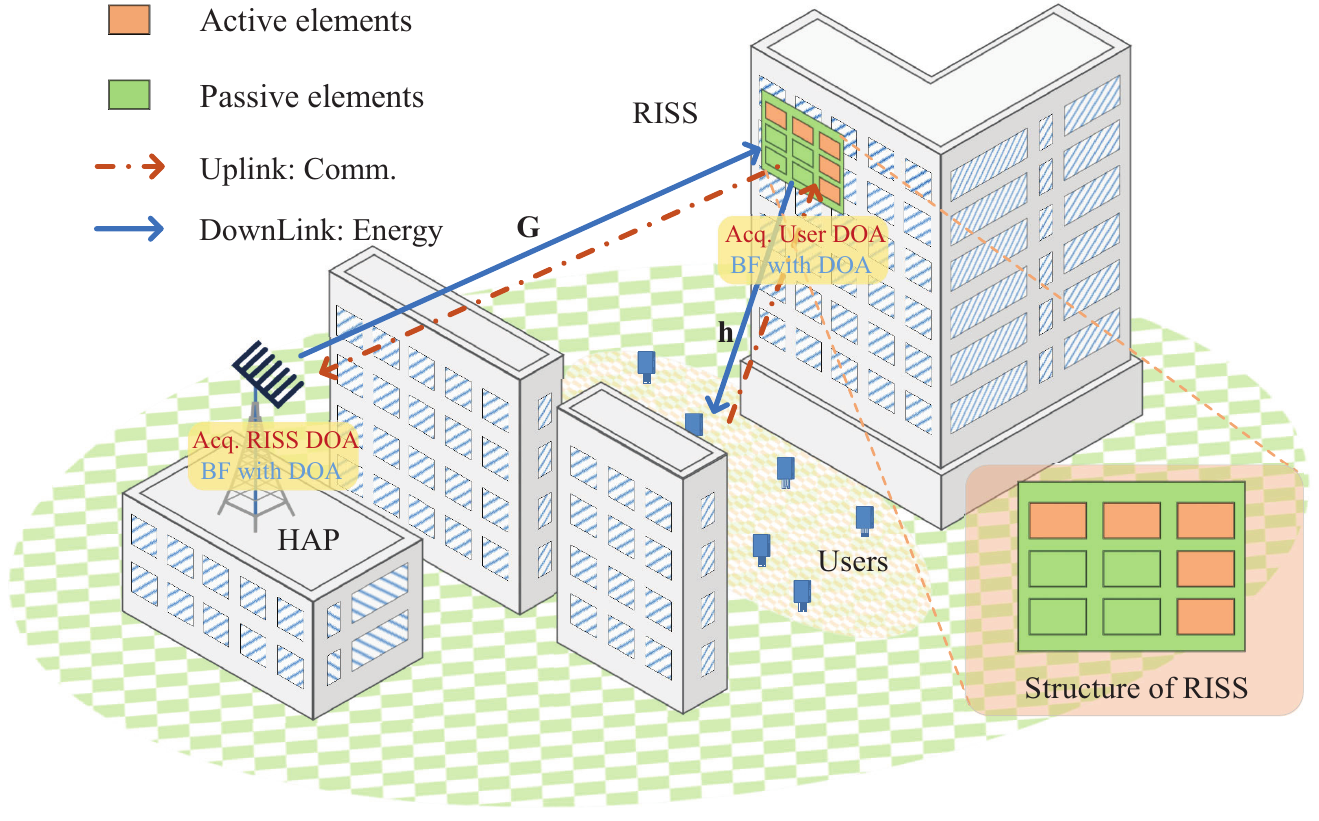}
        \setlength{\abovecaptionskip}{0pt}
        \setlength{\belowcaptionskip}{0pt} 
        \caption{System model of proposed RISS scheme. The RISS is composed of both active and passive elements, which enables the RISS to perform two tasks simultaneously.}
        \label{fig:systemmodel}
    \end{figure}
    In this section, we illustrate the system model of RISS assisted WPCN as well as the transmission protocol for uplink and downlink operations.

    As illustrated in Fig. \ref{fig:systemmodel}, a hybrid access point (HAP) equipped with $M$ antennas as a uniform linear array (ULA), and supports WIT/WET to $K$ single antenna users with the assistance of a RIS composed of $(N_a+N)$ elements ($N_a\ll N$), where $N_a$ represents the number of active elements and $N=N_x\times N_y$ represents the number of passive elements. This RIS, comprising both active and passive elements, serves a dual function of enabling both DOA estimation and WET/WIT enhancement, and henceforth will be referred to as RISS throughout this paper. It is worth mentioning that due to the presence of blockages, there is no direct link between the HAP and the users.
    \subsection{Channel Model}\label{sec:sec2A}
    Quasi-static block fading Rician channels are taken into account. Specifically, the channels between HAP-to-RISS and RISS-to-$k$-th user are denoted as $\mathbf{G}\in\mathbb{C}^{N\times M}$ and $\mathbf{h}_k\in\mathbb{C}^{N\times 1}$, respectively. We use $\kappa_G$ and $\kappa_{h,k}$ to represent the Rician factor of $\mathbf{G}$ and $\mathbf{h}_k$, then we have
    \begin{align}
        &\mathbf{G}=\sqrt{\frac{\kappa_G}{1+\kappa_G}}\bar{\mathbf{G}}+\sqrt{\frac{1}{1+\kappa_G}}\hat{\mathbf{G}},\\
        &\mathbf{h}_k=\sqrt{\frac{\kappa_{h,k}}{1+\kappa_{h,k}}}\bar{\mathbf{h}}_k+\sqrt{\frac{1}{1+\kappa_{h,k}}}\hat{\mathbf{h}}_k,
    \end{align}
    where $\bar{\mathbf{G}}$ ($\bar{\mathbf{h}}_k$) denotes the LoS component and $\hat{\mathbf{G}}$ ($\hat{\mathbf{h}}_k$) denotes the NLoS component of channel $\mathbf{G}$ ($\mathbf{h}_{k}$). $\hat{\mathbf{G}}$ and $\hat{\mathbf{h}}_k$ are both circularly symmetric complex white gaussian (CSCG) and we have $[\hat{\mathbf{G}}]_{i,j}\sim\mathcal{CN}(0, 1),i\in N,j\in M$ and $[\hat{\mathbf{h}}_k]_i\sim\mathcal{CN}(0, 1),i\in N$. Note that the NLoS channel model (i.e., i.i.d. Rayleigh channel) in this paper is a general assumption of the RISS channel\cite{jointlydesign1,jointlydesign2,discretephase1,swipt2,pilotoverhead1,wpcn,sensingirs_dll}, and recent research\cite{correlatedRayleigh} has indicated that Rayleigh channels involving RIS exhibit correlations. We leave the exploration of research pertaining to correlated Rayleigh channels for future investigations.

    Since the RISS can be considered as a uniform planer array (UPA), whereas HAP is a ULA, the LoS component $\bar{\mathbf{G}}$ and $\bar{\mathbf{h}}_k$ under the far-field assumption can be expressed as
    \begin{align}
        &\bar{\mathbf{G}} = \sqrt{MN}\boldsymbol{\alpha}(u_G, v_G)\boldsymbol{\beta}^\mathrm{H}(z_G), \\
        &\bar{\mathbf{h}}_{k} = \sqrt{N}\boldsymbol{\alpha}(u_{h,k}, v_{h,k}), \\
       & \left[\boldsymbol{\alpha}(u, v)\right]_n =  \frac{1}{\sqrt{N_y}}e^{(\text{mod}(n,N_y)-1)\mathbbm{i}v}\frac{1}{\sqrt{N_x}}e^{(\lfloor n/N_x+1 \rfloor-1)\mathbbm{i}u},\label{eqn:vectoralpha}\\
       & \left[\boldsymbol{\beta}(z)\right]_n = \frac{1}{\sqrt{M}}e^{(n-1)\mathbbm{i}z},
    \end{align}
    where $(u,v,z)$ are the spatial phase differences between two adjacent antennas along different axis. And $u=2\pi d\cos(\varphi)/\lambda=\pi\cos(\varphi)$ by setting $d/\lambda=1/2$ without loss of generality, with $d$ and $\lambda$ are the element spacing and carrier wavelength. Similarly, we have $v = \pi\sin(\varphi)\cos(\vartheta)$, $z=\pi\sin(\varpi)$. $\text{mod}(\cdot)$ denotes the remainder operation and $\lfloor \cdot\rfloor$ stands for rounding down operation. And $\varphi_G$ ($\vartheta_G$) and $\varpi_G$ denote the azimuth (elevation) angle of arrival (AOA) and the angle of departure (AOD) from HAP-to-RISS, $\varphi_{h,k}$ ($\vartheta_{h,k}$) denote the azimuth (elevation) angle of departure (AOD) from RISS-to-$k$-th user, respectively.
    \subsection{Transmission Protocols and Elements Placement}\label{sec:IIB}
    \begin{figure}
        \centering
        \includegraphics[width=0.8\linewidth]{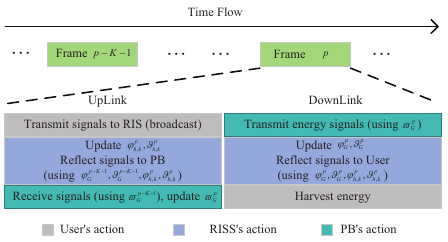}
        \setlength{\abovecaptionskip}{0pt}
        \setlength{\belowcaptionskip}{0pt} 
        \caption{Proposed transmission protocols. }
        \label{fig:transmissionprotocols}
    \end{figure}
    This paper investigates WPCNs that are optimal for WET and WIT scenarios involving users with periodic operation. Specifically, this network employs WIT from a user to the HAP in the uplink, while WET takes place in the downlink to compensate for the consumed energy for constant monitoring of the environment and subsequent uplink transmission.

    We aim to leverage the sensing capabilities of the RISS to further exploit the additional information carried by the signals, specifically the DOA information, as a substitute for conventional channel estimation. This approach aims to achieve a reduction in pilot overhead for channel estimation and decoupling between the RISS and the HAP.
    
    Our study operates under the assumption that each user is allocated a dedicated time frame for WIT and WET, and there is no time overlap between frames\footnote{Considering the scenario of multiple users and frequency division duplex (FDD) system requires more complex DOA estimation, information matching and hardware design, we choose to postpone these aspects to the future work.} as shown in Fig. \ref{fig:transmissionprotocols}, which also means that we can ignore subscript $k$ both of $\mathbf{h}_k$, $\kappa_{h,k}$ and other user-related parameters. Each frame is partitioned into uplink and downlink. During the uplink, the user broadcast signals to the RISS. Then the passive elements of the RISS utilize the AOA of the user (i.e., $\varphi_{h,k}^{p}, \vartheta_{h,k}^{p}$), obtained from the active elements' estimations, along with the slowly varying AOA/AOD of the HAP acquired in the previous frame(i.e., $\varphi_G^{p-K-1}, \vartheta_G^{p-K-1}$), to reflect the signals towards the HAP.\footnote{It is worth noting that rapid algorithms for DOA estimation have been extensively studied, such as using a single snapshot for DOA estimation\cite{singlesnapshot1,singlesnapshot2}. Therefore, in this paper, we consider the time required for sensing to be negligible.} Subsequently, the HAP receives signals with the assistance of AOA $\varpi_G^{p-K-1}$ and updates it to $\varpi_G^{p}$. Similar to uplink, HAP transfers energy to the RISS using $\varpi_G^{p}$, and the RISS updates AOA/AOD (i.e., $\varphi_G^{p}, \vartheta_G^{p}$) accordingly and reflect the signals to the user using fully updated AOA/AOD (i.e., $\varphi_G^{p}, \vartheta_G^{p}, \varphi_{h,k}^{p}, \vartheta_{h,k}^{p}$). It is noteworthy that in both the uplink and downlink, HAP only knows the $\varpi_G$, and RISS only knows $\varphi_G, \vartheta_G, \varphi_{h,k}$ and $\vartheta_{h,k}$. Moreover, there is reciprocity in angle when ignoring the propagation behavior of electromagnetic waves\cite{angleRecip}, which means that the AOA in the uplink is the AOD in the downlink for both RISS and HAP.

    Based on the transmission protocol described above, we suggest performing uplink WIT before downlink WET. This is because at the beginning, the user's DOA information may become outdated, and using them for downlink WET may cause the beams of the RISS to misalign with the user. Alternatively, since the single-antenna user broadcasts signals to the RISS, and the positions of the HAP and the RISS are relatively fixed, it is possible to prioritize the update of the user's DOA information through uplink WIT. This ensures the performance of downlink WET while minimizing any potential sacrifice in WIT performance.
    
    The placement of both active and passive elements of the RISS plays a vital role in the proposed framework. In this paper, the active elements are positioned in an L-array on one side of the RISS, as depicted in Fig. \ref{fig:systemmodel}. The reasons are summarized as follows:
    \begin{enumerate*} 
        \item The algorithms of DOA estimation based on L-array have been extensively studied.
        \item The passive elements within the RISS still form a complete UPA. 
        \item The spatial separation between active and passive elements in the arrangement may potentially mitigates coupling issues in the hardware.
    \end{enumerate*}

    It is noteworthy that the Rician factors and path loss of RISS-to-user and HAP-to-RISS channels are assumed to be known in advance, as these parameters can be determined directly when users are deployed. This means that the RISS can design its phase shift based on the DOA estimation of active elements, without requiring a feedback link between the RISS controller and the HAP. Such a self-contained operation enables the RISS to function independently, rendering it more amenable for deployment and migration\footnote{For the case where the parameters are unknown, we can also use active elements of RISS to estimate Rician factors\cite{RicianFactorEst2,RicianFactorEst3} and path loss directly.}.
    \section{Performance Analysis of Proposed scheme}\label{sec:sec3}
    This section focuses on multiple-in-single-out (MISO) system. Firstly, we present the design of transmitter precoders and RISS reflecting phases, aiming to evaluate the average performance of energy reception and communication. Then we also investigate the impact of DOA estimation errors, which are modeled with an L-array, and provide a comprehensive analysis of its effect on system performance, along with a closed-form expression of energy reception. Finally, we further analyzed the statistical characteristics of the energy at the receiver through moment-matching, making it possible to meet the outage probability requirements of the receiver for Rician channel.
    \subsection{Energy reception and Communication Performance}\label{sec:EnergyComm}
    With the assistance of RISS, the average energy received by the $k$-th user can be expressed as
    \begin{align}
        \mathbb{E}&\{E_k\}\nonumber\\
        =&\mathbb{E}\left\{\left|\sqrt{\varrho_{R2U,k}\varrho_{H2R}}\mathbf{h}_k^\mathrm{H}\boldsymbol{\Theta}\mathbf{G}\mathbf{w}_ks_k\right|^2\right\}\nonumber\\
        =&\varrho_{H2U,k}\mathbb{E}\left\{\left|\left(\sqrt{\frac{\kappa_{h,k}}{1+\kappa_{h,k}}}\bar{\mathbf{h}}_k^\mathrm{H}+\sqrt{\frac{1}{1+\kappa_{h,k}}}\hat{\mathbf{h}}_k^\mathrm{H}\right)\mathrm{diag}\{\theta_{1,{h}},\cdots,\theta_{N,{h}}\}\right.\right.\nonumber\\
        &\left.\left.\times\mathrm{diag}\{\theta_{1,{G}},\cdots,\theta_{N,{G}}\}\left(\sqrt{\frac{\kappa_{G}}{1+\kappa_{G}}}\bar{\mathbf{G}}+\sqrt{\frac{1}{1+\kappa_{G}}}\hat{\mathbf{G}}\right)\mathbf{w}_k\right|^2\right\}\label{eqn:Ek_org},
    \end{align}
    where $s_k$ is the normalized signal, i.e., $\mathbb{E}\{s_k^\mathrm{H}s_k\}=1$. $\mathrm{diag}\{\theta_{1,{h}},\cdots,\theta_{N,{h}}\}\cdot\mathrm{diag}\{\theta_{1,{G}},\cdots,\theta_{N,{G}}\}\cdot\mathrm{diag}\{\beta_1,\cdots,\beta_N\}=\boldsymbol{\Theta}\in\mathbb{C}^{N\times N}$ is the diagonal reflection matrix of RISS, where $\{\theta_{i,{h}}, \theta_{i,{G}}\}$ and $\beta_i\in[0,1], \forall i\in N$ are the phase shift and amplitude
    reflection coefficients, respectively. We set $\beta_i=1,\forall i\in N$ to maximize the reflection signal power without loss of generality. $\varrho_{R2U,k}$ and $\varrho_{H2R}$ denote the path loss of RISS-to-$k$-th user and HAP-to-RISS, respectively, and $\varrho_{H2U,k}=\varrho_{R2U,k}\varrho_{H2R}$ denotes the total cascade path loss of HAP-to-$k$-th user. $\mathbf{w}_k$ denotes the precoder for the $k$-th user.

    We first focus on the LoS component of channel $\mathbf{G}$ since we can obtain the AOA/AOD information from DOA estimation, which is expressed as
    \begin{align}
        &\sqrt{\frac{\kappa_G}{1+\kappa_G}}\mathrm{diag}\{\theta_{1,{G}},\cdots,\theta_{N,{G}}\}\bar{\mathbf{G}}\mathbf{w}_k\nonumber\\
        &=\sqrt{\frac{MN\kappa_G}{1+\kappa_G}} \mathrm{diag}\{\theta_{1,{G}},\cdots,\theta_{N,{G}}\}\boldsymbol{\alpha}(u_G, v_G)\boldsymbol{\beta}^\mathrm{H}(z_G) \mathbf{w}_k,
    \end{align}
    where $\mathbf{w}_k$ represents the transmission precoder at HAP. Since HAP only knows $\varpi_G$ (corresponding to $z_G$), while RISS has knowledge of $u_G$ and $v_G$, we can use the RISS phase shift matrix $\mathrm{diag}\{\theta_{1,{G}},\cdots,\theta_{N,{G}}\}$ to maximize $\mathrm{diag}\{\theta_{1,{G}},\cdots,\theta_{N,{G}}\}\boldsymbol{\alpha}(u_G, v_G)$. Similarly, we may adopt the transmission precoder $\mathbf{w}_k$ to maximize $\boldsymbol{\beta}^\mathrm{H}(z_G)\mathbf{w}_k$, which leads to the determination of $\mathbf{w}_k$ as the maximal ratio transmission (MRT) (i.e., $\mathbf{w}_k=\sqrt{P_E}\frac{\boldsymbol{\beta}(z_G)}{||\boldsymbol{\beta}(z_G)||}$, where $P_E$ is the transmit power for WET). Performing a similar analysis on $\mathbf{h}_k$, we have
     \begin{align}
        &\bar{\mathbf{h}}_k^\mathrm{H}\mathrm{diag}\{\theta_{1,{h}},\cdots,\theta_{N,{h}}\}=\mathbf{1}_{N\times 1}^\mathrm{T},\nonumber\\
        &\mathrm{diag}\{\theta_{1,{G}},\cdots,\theta_{N,{G}}\}\bar{\mathbf{G}}\mathbf{w}_k=\mathbf{1}_{N\times 1} \sqrt{MP_E}\label{eqn:hG_cons},
    \end{align}
    where the optimal phase shift coefficients are $\theta_{i,{h}}=e^{(\text{mod}(i,N_y)-1)\mathbbm{i}v_h+(\lfloor i/N_x+1 \rfloor-1)\mathbbm{i}u_h}, i\in N$, $\theta_{i,{G}}=e^{-(\text{mod}(i,N_y)-1)\mathbbm{i}v_G-(\lfloor i/N_x+1 \rfloor-1)\mathbbm{i}u_G}, i\in N$, $u_h = \pi\cos(\varphi_{h,k}),v_h=\pi\sin(\varphi_{h,k})\cos(\vartheta_{h,k})$ and $u_G = \pi\cos(\varphi_{G}),v_G=\pi\sin(\varphi_{G})\cos(\vartheta_{G})$. Eq. \eqref{eqn:hG_cons} implies that the LoS components become constant terms. And for NLoS components, we have
    \begin{align}
        &\hat{\mathbf{h}}_k^\S=\hat{\mathbf{h}}_k^\mathrm{H}\mathrm{diag}\{\theta_{1,{h}},\cdots,\theta_{N,{h}}\}\in\mathbb{C}^{1\times N}\sim\mathcal{CN}(0,\mathbf{I}_{N}),\label{eqn:h_S}\\
        &\hat{\mathbf{G}}^\S=\frac{\mathrm{diag}\{\theta_{1,{G}},\cdots,\theta_{N,{G}}\}\hat{\mathbf{G}}\mathbf{w}_k}{\sqrt{P_E}}\in\mathbb{C}^{N\times 1}\sim\mathcal{CN}(0,\mathbf{I}_{N}),\label{eqn:G_S}
    \end{align}
    \begin{table}[]
        \centering
        \caption{The performance ratio between the MISO and SISO system various with the number of RISS passive elements and Rician factor.}
        \begin{tabular}{c|ccccc}
        \hline
        \textbf{Cond.} &  $N\to \infty$&  $\kappa_{h,k}\to \infty$& $\kappa_{h,k}\to 0$ & $\kappa_G\to \infty$ & $\kappa_G\to 0$ \\
        \hline
        \textbf{Ratio}& $M$ & $\frac{NMk_{G}+1}{Nk_{G}+1}\leq M$ & $\frac{Mk_{G}+1}{k_{G}+1}\leq M$ & $M$ &  $1$\\
        \hline
        \end{tabular} \label{table:condratio}
    \end{table}
    since CSCG is circularly symmetric and the precoder $\mathbf{w}_k$ aggregates $\hat{\mathbf{G}}$ in the column, resulting in an invariant variance. Finally we can rewrite Eq. \eqref{eqn:Ek_org} as Eq. \eqref{eqn:cons_sta} at the top of this page.
    \begin{figure*}
        \begin{align}
            &\mathbb{E}\left\{E_k\right\}=\varrho_{H2U,k}P_E\mathbb{E}\left\{\left|\left(\sqrt{\frac{\kappa_{h,k}}{1+\kappa_{h,k}}}{\mathbf{1}}^\mathrm{T}_{N\times 1}+\sqrt{\frac{1}{1+\kappa_{h,k}}}\hat{\mathbf{h}}_k^\S\right)\left(\sqrt{\frac{\kappa_{G}}{1+\kappa_{G}}}{\mathbf{1}}_{N\times 1}\sqrt{M}+\sqrt{\frac{1}{1+\kappa_{G}}}\hat{\mathbf{G}}^\S\right)\right|^2\right\} \nonumber\\
            &=\frac{\varrho_{H2U,k}P_E}{(1+\kappa_{h,k})(1+\kappa_{G})}\mathbb{E}\left\{\left|N\sqrt{M}\sqrt{\kappa_{h,k}\kappa_{G}}+\sqrt{\kappa_{G}}\hat{\mathbf{h}}_k^\S\mathbf{1}_{N\times 1}\sqrt{M}+\sqrt{\kappa_{h,k}}\mathbf{1}_{N\times 1}^\mathrm{T}\hat{\mathbf{G}}^\S+\hat{\mathbf{h}}_k^\S\hat{\mathbf{G}}^\S\right|^2\right\}\nonumber\\
            &\overset{(a)}{=}\frac{\varrho_{H2U,k}P_E}{(1+\kappa_{h,k})(1+\kappa_{G})}\mathbb{E}\left\{\underbrace{N^2M\kappa_{h,k}\kappa_{G}}_{\textbf{constant}}+\underbrace{M\kappa_{G}\hat{\mathbf{h}}_k^\S\mathbf{1}_{N\times 1}\mathbf{1}_{N\times 1}^\mathrm{T}\hat{\mathbf{h}}_k^{\S, \mathrm{H}}+\kappa_{h,k}\mathbf{1}_{N\times 1}^\mathrm{T}\hat{\mathbf{G}}^\S\hat{\mathbf{G}}^{\S,\mathrm{H}}\mathbf{1}_{N\times 1}+\hat{\mathbf{h}}_k^\S\hat{\mathbf{G}}^\S\hat{\mathbf{G}}^{\S,\mathrm{H}}\hat{\mathbf{h}}_k^{\S, \mathrm{H}}}_{\textbf{statistic}}\right\}\nonumber\\
            &=\frac{\varrho_{H2U,k}P_E}{(1+\kappa_{h,k})(1+\kappa_{G})}\left(N^2M\kappa_{h,k}\kappa_{G}+NM\kappa_{G}+N\kappa_{h,k}+N\right)\label{eqn:cons_sta}.
    \end{align}
    \hrulefill
    \end{figure*}
    Where $(a)$ of Eq. \eqref{eqn:cons_sta} can be attributed to the fact that a non-zero expectation value arises only when the CSCG matrix is multiplied by its own transpose. Furthermore, the single-in-single-out (SISO) system can be obtained by setting $M=1$, and we record it as
    \begin{align}
        \mathbb{E}&\left\{E_{k, \text{SISO}}\right\} \nonumber\\
        &= \frac{\varrho_{H2U,k}P_E}{(1+\kappa_{h,k})(1+\kappa_{G})}\left(N^2\kappa_{h,k}\kappa_{G}+N\kappa_{G}+N\kappa_{h,k}+N\right).
    \end{align}
    Thus, the relationship between MISO and SISO system is
    \begin{align}
        \frac{\mathbb{E}\left\{E_{k,\text{MISO}}\right\}}{\mathbb{E}\left\{E_{k,\text{SISO}}\right\}} &= \frac{\mathbb{E}\left\{E_{k}\right\}}{\mathbb{E}\left\{E_{k,\text{SISO}}\right\}}\nonumber\\&=\frac{N^2M\kappa_{h,k}\kappa_{G}+NM\kappa_{G}+N\kappa_{h,k}+N}{N^2\kappa_{h,k}\kappa_{G}+N\kappa_{G}+N\kappa_{h,k}+N}.\label{eqn:performanceRatio}
    \end{align}
  
    \begin{Rem}
        The received energy performance shown in Eq. \eqref{eqn:cons_sta} demonstrates that the constant component increases with $N^2$, while statistic components rise with $N$. Note that the statistic components increase slower than the constant counterpart in our proposed scheme. Moreover, the performance improvement of the MISO system compared to the SISO system, as shown in Table II, requires non-zero $\kappa_G$ to fully utilize the benefits of multiple antennas HAP, for example, careful selection of RISS placement to achieve this necessary requirement\footnote{In fact, $\kappa_G$ represents the power ratio between the LoS component to the diffusive component in the HAP to RISS channel $\mathbf{G}$. When deploying a RISS system, the positions of HAP and RISS are often fixed, while the positions of users may be random and mobile. This implies that we may ensure $\kappa_G>0$ through certain deployment strategies, such as placing RISS at elevated locations to reduce the power of the NLoS components caused by reflections, building obstructions, and other factors.}. Furthermore, it is worth noting that we retain the subscript $k$ to illustrate performance variations with different $\mathbf{h}_k$. Observe from Eq. \eqref{eqn:cons_sta} that the mean received energy for different users is primarily influenced by the path loss $\varrho_{R2U,k}$ from RISS to the user and Rician factor $\kappa_{h,k}$. While the randomness and mobility of user positions lead to variations in the value of $\kappa_{h,k}$, as long as $\kappa_G>0$, we can always benefit from multiple antennas.
        \label{remark:1}
    \end{Rem}

    To maintain the clarity, we formally remove the subscript $k$, as we have elucidated in Section \ref{sec:IIB}. For WIT, we utilize maximal ratio combining (MRC) denoted by $\tilde{\mathbf{w}}, ||\tilde{\mathbf{w}}||^2=1$ in HAP, while keeping all other parameters unchanged in the uplink. Then the signal-to-noise ratio (SNR) can be expressed as
    \begin{align}
        SNR = \frac{\varrho_{U2H}P_I\left|\tilde{\mathbf{G}}^\mathrm{H}\tilde{\boldsymbol{\Theta}}\tilde{\mathbf{h}}\tilde{\mathbf{w}}\right|^2}{\sigma^2},\label{eqn:snr_ori}
    \end{align}
    where $\sigma^2$ is the noise power, $P_I$ is the transmit power for WIT, $\tilde{\mathbf{G}}\in\mathbb{C}^{N\times M}$, $\tilde{\boldsymbol{\Theta}}\in\mathbb{C}^{N\times N}$ and $\tilde{\mathbf{h}}\in \mathbb{C}^{N\times 1}$ denote user-to-RISS channel, diagonal phase shift matrix of RISS in uplink and RISS-to-HAP channel, respectively. We note that there is no interference signals since each user is allocated a dedicated time frame for WIT. Additionally, given the reciprocity in angles, as mentioned in Section \ref{sec:IIB}, and with $\varrho_{U2H}=\varrho_{H2U}$, it can be inferred that the average performance of uplink and downlink is equivalent for the entire system. The upper bound of ergodic spectral efficiency can be expressed through Jensen's inequality (the tight closed-form solution is given in Lemma \ref{lemma:lemma4}) as
    \begin{align}
        C_{MISO}&=\mathbb{E}\{\log_2(1+SNR)\}\leq\log_2(1+\mathbb{E}\{SNR\})\nonumber\\
        &=\log_2\left(1+\frac{N^2M\kappa_h\kappa_G+NM\kappa_G+N\kappa_h+N}{(\varrho_{H2U}P_I)^{-1}(1+\kappa_h)(1+\kappa_G)\sigma^2}\right).\label{eqn:upperbound_cc}
    \end{align}

    Up to this point, our discussions have revolved around the assumption of accurate angle estimation. However, it is crucial to address the issue of imperfect angle estimation in this paper. In the forthcoming section, we will delve into the impacts of angle estimation errors on the overall system.

    \subsection{DOA Errors Analysis}
    \begin{figure}
        \centering
        \includegraphics[width=0.9\linewidth]{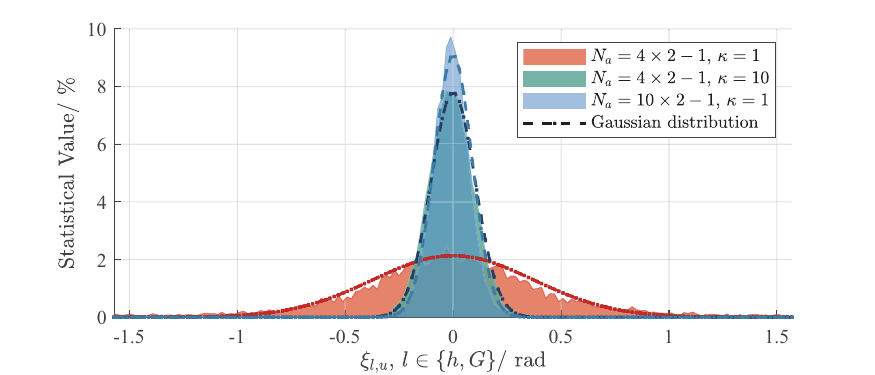}
        \setlength{\abovecaptionskip}{0pt}
        \setlength{\belowcaptionskip}{0pt} 
        \caption{We employ the ROOT-MUSIC algorithm, utilizing an L-array consisting of $N_a=7$ or $19$ active elements, to estimate the two-dimensional DOA within the range of $[-\pi/3, \pi/3]$. The Rician factor is considered as either 1 or 10. The obtained error statistics of the spatial phase differences (i.e., $\xi_{h,u}$ or $\xi_{G,u}$ as given in Eq. \eqref{eqn:uvzerror}) are illustrated above, based on Monte Carlo simulations.}
        \label{fig:doaerror}
    \end{figure}

    We have discussed the WET and WIT performance in the scenario of accurate angle estimation in Section \ref{sec:EnergyComm}. Nevertheless, in practical DOA estimation, the accuracy is affected by various factors such as the number of active elements, the value of the Rician factor, receiver distortions, sensor location errors, mutual coupling errors, etc. \cite{VariousErrors}. Hence, in this section, we aim to investigate the impact of DOA estimation errors on the system's performance.

   Indeed, many studies employ Gaussian distribution to model DOA estimation errors \cite{gaussianerror, gaussianerror2}. We then plot both the actual probability density function (PDF) of the Gaussian distribution and the statistical distribution of DOA estimation errors in the spatial phase differences in Fig. \ref{fig:doaerror}. Observe from Fig. \ref{fig:doaerror} that the errors in the spatial phase differences approximately follows a Gaussian distribution. However, due to the presence of sine and cosine terms, deriving a closed-form expression for the distribution of angles $(\varphi,\vartheta,\varpi)$ is challenging. To address this issue, we assume that the errors in the angles $(\varphi,\vartheta,\varpi)$ also follow Gaussian distributions. Subsequently, we employ a fitting technique to establish a relationship between the Gaussian errors in the spatial phase differences $(u,v,z)$ and the angle errors, thereby obtaining a closed-form expression.

    We first modeled the DOA estimation errors of MISO system in spatial phase differences $(u,v,z)$ as
    \begin{align}
        &u_h-u_h^e=\xi_{h,u}\sim\mathcal{N}(0,{\sigma^2_{h,u}}),\nonumber\\
        &v_h-v_h^e=\xi_{h,v}\sim\mathcal{N}(0,{\sigma^2_{h,v}}),\nonumber\\
        &u_G-u_G^e=\xi_{G,u}\sim\mathcal{N}(0,{\sigma^2_{G,u}}),\nonumber\\
        &v_G-v_G^e=\xi_{G,v}\sim\mathcal{N}(0,{\sigma^2_{G,v}}),\nonumber\\
        &z_G-z_G^e = \xi_{G,z}\sim\mathcal{N}(0,{\sigma^2_{G,z}}),\label{eqn:uvzerror}
    \end{align}
    where $u_h,v_h,u_G,v_G,z_G$ denote the accurate spatial phase differences, and $u^e_h,v^e_h,u^e_G,v^e_G,z^e_G$ denote the practical spatial phase differences calculated from DOA estimation. We can rewrite Eq. \eqref{eqn:hG_cons} as
    \begin{align}
        &\bar{\mathbf{h}}^\mathrm{H}\mathrm{diag}\{\theta_{1,{h}},\cdots,\theta_{N,{h}}\}=\sqrt{N}\boldsymbol{\alpha}^\mathrm{H}(\xi_{h,u},\xi_{h,v})\label{eqn:not_cons_h},\\
        &\frac{\mathrm{diag}\{\theta_{1,{G}},\cdots,\theta_{N,{G}}\}\bar{\mathbf{G}}\mathbf{w}}{\sqrt{P_E}}=\sqrt{N}\boldsymbol{\alpha}(\xi_{G,u},\xi_{G,v})\mathbf{1}^\mathrm{T}_{M\times 1}\boldsymbol{\beta}(\xi_{G,z}).\label{eqn:not_cons_G}
    \end{align}
    \begin{IEEEproof}
        Please refer to Appendix \ref{app:A} for detailed proof.
    \end{IEEEproof}

    Since the characteristics of the last two terms of statistical components have not changed in Eq. \eqref{eqn:cons_sta}, we first focus on the change of constant component. Replace Eq. \eqref{eqn:hG_cons} with Eq. \eqref{eqn:not_cons_h} and Eq. \eqref{eqn:not_cons_G}, the constant component of Eq. \eqref{eqn:cons_sta} can be rewritten as
    \begin{align}
        &N^2M\kappa_h\kappa_G\to\nonumber\\
        &\kappa_h\kappa_G\left|\sqrt{N}\boldsymbol{\alpha}^\mathrm{H}(\xi_{h,u},\xi_{h,v})\sqrt{N}\boldsymbol{\alpha}(\xi_{G,u},\xi_{G,v})\mathbf{1}^\mathrm{T}_{M\times 1}\boldsymbol{\beta}(\xi_{G,z})\right|^2\nonumber\\
        =&\kappa_h\kappa_G\left|\sqrt{N}\mathbf{1}_{N\times 1}^\mathrm{T}\boldsymbol{\alpha}(\xi_{h,u}+\xi_{G,u},\xi_{h,v}+\xi_{G,v})\mathbf{1}^\mathrm{T}_{M\times 1}\boldsymbol{\beta}(\xi_{G,z})\right|^2.\label{eqn:error_cons_1}
    \end{align}
    \begin{Lemma}\label{lemma:lemma1}
        The mathematical expectation of Eq. \eqref{eqn:error_cons_1} can be expressed as
        \begin{align}
            \kappa_h\kappa_G\sum_{i=0}^{N_x-1}\sum_{k=0}^{N_x-1}e^{-\frac{(i-k)^2\left(\sigma_{h,u}^2+\sigma_{G,u}^2\right)}{2}}&\sum_{i=0}^{N_y-1}\sum_{k=0}^{N_y-1}e^{-\frac{(i-k)^2\left(\sigma_{h,v}^2+\sigma_{G,v}^2\right)}{2}}\nonumber\\
            &\times \frac{1}{M}\sum_{i=0}^{M-1}\sum_{k=0}^{M-1}e^{-\frac{(i-k)^2\sigma_{G,z}^2}{2}}.\label{eqn:error_cons}
        \end{align}
    \end{Lemma}
    \begin{IEEEproof}
        Please refer to Appendix \ref{app:B} for detailed proof.
    \end{IEEEproof}

    According to Lemma \ref{lemma:lemma1}, we can rewrite the mathematical expectation of received energy (which corresponds to Eq. \eqref{eqn:cons_sta}) as
    \begin{align}
        \mathbb{E}\left\{E_k\right\}=&\frac{\varrho_{H2U}P_E}{(1+\kappa_{h})(1+\kappa_{G})}\left(\kappa_h\kappa_G\sum_{i=0}^{N_x-1}\sum_{k=0}^{N_x-1}e^{-\frac{(i-k)^2\left(\sigma_{h,u}^2+\sigma_{G,u}^2\right)}{2}}\right.\nonumber\\
        &\times \sum_{i=0}^{N_y-1}\sum_{k=0}^{N_y-1}e^{-\frac{(i-k)^2\left(\sigma_{h,v}^2+\sigma_{G,v}^2\right)}{2}}\frac{1}{M}\sum_{i=0}^{M-1}\sum_{k=0}^{M-1}e^{-\frac{(i-k)^2\sigma_{G,z}^2}{2}}\nonumber\\
        &\qquad\left.+\frac{1}{M}\sum_{i=0}^{M-1}\sum_{k=0}^{M-1}e^{-\frac{(i-k)^2\sigma_{G,z}^2}{2}}N\kappa_{G}+N\kappa_{h}+N\right).\label{eqn:erroronUVZ}
    \end{align}

    Up to now, we finished the analysis of error model in the spatial phase differences $(u, v, z)$. It is worth emphasizing the significance of the analysis conducted on spatial phase differences in Eq. \eqref{eqn:erroronUVZ}. This analysis provides a universally applicable, and rigorously derived closed-form solution that can be employed across a wide spectrum of scenarios\cite{VariousErrors,gaussianerror,gaussianerror2}. And then we can get the error model in angles as\footnote{Due to the fact that DOA errors are primarily associated with channel conditions and array structure, and the elevation angle and azimuth angle are estimated using the same array when a signal is received, it is posited that the elevation angle and azimuth angle error (e.g., $\vartheta_G-\vartheta_{DOA,G}$ and $\varphi_G-\varphi_{DOA, G}$) share the same variance. However, the variances of $\vartheta_G-\vartheta_{DOA,G}$ and $\vartheta_h-\vartheta_{DOA,h}$ are different.}
    \begin{align}
        &\vartheta_G-\vartheta_{DOA,G} \sim\mathcal{N}(0, \sigma^2_{DOA, G}),\nonumber\\
        &\varphi_G-\varphi_{DOA,G} \sim\mathcal{N}(0, \sigma^2_{DOA, G}),\nonumber
        \\
        &\varpi_G-\varpi_{DOA,G} \sim\mathcal{N}(0, \sigma^2_{DOA, P}),\nonumber\\
        &\vartheta_h-\vartheta_{DOA,h} \sim\mathcal{N}(0, \sigma^2_{DOA, h}),\nonumber\\
        &\varphi_h-\varphi_{DOA,h} \sim\mathcal{N}(0, \sigma^2_{DOA, h}).
    \end{align}

    The fitting technique yields the definitive equation for the average energy receiving performance under DOA estimation errors as follows
    
    \begin{align}
        \mathbb{E}&\left\{E_k^{DOA}\right\}\nonumber\\
        =&\frac{\varrho_{H2U}P_E}{(1+\kappa_{h})(1+\kappa_{G})}\left(\kappa_h\kappa_G\sum_{i=0}^{N_x-1}\sum_{k=0}^{N_x-1}e^{-\eta_u(i-k)^2\left(\sigma_{DOA,h}^2+\sigma_{DOA,G}^2\right)}\right.\nonumber\\
        &\times  \sum_{i=0}^{N_y-1}\sum_{k=0}^{N_y-1}e^{-\eta_v(i-k)^2\left(\sigma_{DOA,h}^2+\sigma_{DOA,G}^2\right)}\times\frac{1}{M}\sum_{i=0}^{M-1}\sum_{k=0}^{M-1}e^{-\eta_z(i-k)^2\sigma_{DOA,P}^2}\nonumber\\
        &\qquad\quad+\left.\frac{1}{M}\sum_{i=0}^{M-1}\sum_{k=0}^{M-1}e^{-\eta_z(i-k)^2\sigma_{DOA,P}^2}N\kappa_{G}+N\kappa_{h}+N\right),\label{eqn:err_final_closeform}
    \end{align}
    where $\eta_u=4.3575, \eta_v=1.395,\eta_z = 2.15$\footnote{We can acquire the results by iterating among the search parameters $\eta_u$, $\eta_v$, and $\eta_z$. In practice, we can begin by fitting the closed-form SISO model and deducing the values of $\eta_u$ and $\eta_v$ through a two-dimensional search. Afterwards, we can employ $\eta_u$ and $\eta_v$ as the corresponding parameters in the MISO model and ascertain the value of $\eta_z$ through a one-dimensional search.}.
    \begin{Lemma}
        Based on Eq. \eqref{eqn:err_final_closeform}, we can also obtain the lower bound and upper bound of receiving energy. Since
        \begin{align}
            &N\leq\sum_{i=0}^{N_x-1}\sum_{k=0}^{N_x-1}e^{-\eta_u(i-k)^2\left(\sigma_{DOA,h}^2+\sigma_{DOA,G}^2\right)}\nonumber\\
            &\qquad\qquad\qquad\qquad\times\sum_{i=0}^{N_y-1}\sum_{k=0}^{N_y-1}e^{-\eta_v(i-k)^2\left(\sigma_{DOA,h}^2+\sigma_{DOA,G}^2\right)}\leq N^2,\nonumber\\
            &1\leq\frac{1}{M}\sum_{i=0}^{M-1}\sum_{k=0}^{M-1}e^{-\eta_z(i-k)^2\sigma_{DOA,P}^2}\leq M,&
        \end{align}
        we have
        \begin{align}
           &\frac{\left(N\kappa_{h}\kappa_{G}+N\kappa_{G}+N\kappa_{h}+N\right)}{(1+\kappa_{h})(1+\kappa_{G})}\nonumber\\
            &\qquad\qquad\leq \frac{\mathbb{E}\left\{E_k^{DOA}\right\}}{\varrho_{H2U}P_E}\leq\nonumber\\
            &\qquad\qquad\qquad\qquad\frac{\left(N^2M\kappa_{h}\kappa_{G}+NM\kappa_{G}+N\kappa_{h}+N\right)}{(1+\kappa_{h})(1+\kappa_{G})}.
        \end{align}
    \end{Lemma}
    \subsection{Outage Probability Analysis}\label{sec:outagePro}
        In the previous content, we examined the utilization of active elements in RISS to facilitate DOA estimation and assist in WET and WIT. However, our analysis primarily focused on the system's average performance without considering the outage probability performance. In this section, we first assume accurate DOA estimation (as discussed in Section \ref{sec:EnergyComm}), and known Rician factors $\kappa_h$ and $\kappa_G$, as well as the cascade path loss $\varrho_{H2U}$, and then expand to imperfect DOA estimation. We employ moment-matching techniques to derive the cumulative distribution function (CDF) of received energy and establish the relationship between transmit power and the user's outage probability.

        Specifically, since the NLoS component of both HAP-to-RISS and RISS-to-user channels are unknown, we still follow the previous configuration as analyzed in Eq. \eqref{eqn:cons_sta}, and the instantaneous received signal can be expressed as
        \begin{align}
            E_{k}=\frac{\varrho_{H2U}P_E}{{(1+\kappa_{h})(1+\kappa_{G})}}&\left|N\sqrt{M}\sqrt{{\kappa_{h}\kappa_{G}}}+\sqrt{{\kappa_{G}}}\hat{\mathbf{h}}^\S\mathbf{1}_{N\times 1}\sqrt{M}\right.\nonumber\\
            &\qquad\left.+\sqrt{{\kappa_{h}}}\mathbf{1}_{N\times 1}^\mathrm{T}\hat{\mathbf{G}}^\S+\hat{\mathbf{h}}^\S\hat{\mathbf{G}}^\S\right|^2.\label{eqn:insE}
        \end{align}
        \begin{figure*}[ht]
            \begin{align}
                &\alpha_{E}=\frac{N\left(NM\kappa_h\kappa_G+M\kappa_G+\kappa_h+1\right)^2}{N\left(\kappa_G^2M^2+\kappa_h^2+1\right)+2\left(\kappa_h\kappa_GMN\left(\kappa_GMN+\kappa_hN+N+5\right)+\left(N+1\right)\left(\kappa_GM+\kappa_h\right)+\left(\kappa_GM+\kappa_h+1\right)\right)},\label{eqn:alpha_E}\\
                &\beta_{E}=\frac{\left(\varrho_{H2U}P_E\right)^{-1}\left(NM\kappa_h\kappa_G+M\kappa_G+\kappa_h+1\right)\left(1+\kappa_h\right)\left(1+\kappa_G\right)}{N\left(\kappa_G^2M^2+\kappa_h^2+1\right)+2\left(\kappa_h\kappa_GMN\left(\kappa_GMN+\kappa_hN+N+5\right)+\left(N+1\right)\left(\kappa_GM+\kappa_h\right)+\left(\kappa_GM+\kappa_h+1\right)\right)},\label{eqn:beta_E}
            \end{align}
            \hrulefill
        \end{figure*}
    It is noteworthy that the product of Gaussian distributions fails to comply with the Gaussian distribution and is referred to as the complex double Gaussian distribution\cite{CNN}. It is challenging to obtain precise results, which has prompted us to employ the moment-matching \cite{momentmatching} method to explicate the distribution of $E_{k}$ in this paper.
    \begin{Lemma}\label{lemma:lemma3}
        The $E_{k}$ obeys the Gamma distribution with the shape parameter $\alpha_{E}$ and rate parameter $\beta_{E}$, where $\alpha_{E}$ and $\beta_{E}$ are given in Eq. \eqref{eqn:alpha_E} and Eq. \eqref{eqn:beta_E} at the top of the next page.
    \end{Lemma}
    \begin{IEEEproof}
        Please refer to Appendix \ref{app:C} for detailed proof.
    \end{IEEEproof}

    According to Lemma \ref{lemma:lemma3}, the transmit power which can satisfy the outage probability $P_{out}$ and energy threshold $T_{thre}$ can be expressed as
    \begin{align}
        P_{trans} = \frac{T_{thre}}{G^{-1}_{{\alpha_E, \beta_E}}(1-P_{out})}.\label{eqn:transpower}
    \end{align}
    Where $G_{\alpha_E, \beta_E}^{-1}(\cdot)$ represents the inverse cumulative distribution function (ICDF) of the Gamma distribution with shape parameter $\alpha_E$ and rate parameter $\beta_E$.

    The adoption of the transmit power specified in Eq. \eqref{eqn:transpower} can ensure the user's outage probability demand is effectively met, thereby significantly enhancing the quality of service (QoS).

    Based on Lemma \ref{lemma:lemma3} and Eq. \eqref{eqn:snr_ori}, we can give a tight closed-form expression of ergodic spectral efficiency in Lemma \ref{lemma:lemma4}.
    \begin{Lemma}\label{lemma:lemma4}
        The ergodic spectral efficiency can be expressed as
        \begin{align}
            C_{MISO}&=\mathbb{E}\{\log_2(1+SNR)\}\nonumber\\
            &=\frac{1}{\Gamma(\alpha_I)\ln 2}G^{1,3}_{3,2} \left(\frac{P_{I}\varrho_{H2U}}{\beta_I\sigma^2} \middle| {^{1-\alpha_I, 1, 1}_{1,\ \ 0}}\right),\label{eqn:closedformcc2}
        \end{align}
        where $\Gamma(\cdot)$ denote the Gamma function, $\alpha_I=\alpha_E$, $\beta_I = \beta_EP_E\varrho_{H2U}$ and $G^{m,n}_{p,q} \left(x \middle| {^{a_1,\cdots,a_n,\cdots,a_p}_{b_1,\cdots,b_m,\cdots,b_q}}\right)$ is Meijer G function.
    \end{Lemma}
    \begin{IEEEproof}
        Denoted $\left|\tilde{\mathbf{G}}^\mathrm{H}\tilde{\boldsymbol{\Theta}}\tilde{\mathbf{h}}\tilde{\mathbf{w}}\right|^2$ by variable $Z$, and according to Lemma \ref{lemma:lemma3}, we have $Z\sim \text{Gamma}(\alpha_I, \beta_I)$, and the PDF of variable Z is
        \begin{align}
            f_{\text{Gamma}, Z}(z)=\frac{\beta_I^{\alpha_I}}{\Gamma(\alpha_I)} z^{\alpha_I-1} e^{-\beta_I z}.
        \end{align}
        Thus Eq. \eqref{eqn:upperbound_cc} can be rewritten as
        \begin{align}
            C_{MISO} =& \mathbb{E}\left\{\log_2\left(1+\frac{\varrho_{H2U}P_I}{\sigma^2}Z\right)\right\}\nonumber\\
            \overset{(a)}{=}&\frac{\sigma^2}{\varrho_{H2U}P_I\ln2}\int_{0}^{\infty}\ln(1+\zeta)f_{\text{Gamma}, Z}\left(\frac{\zeta\sigma^2}{\varrho_{H2U}P_I}\right)\mathrm{d}\zeta\nonumber\\
            \overset{(b)}{=}&\frac{\beta_I^{\alpha_I}\sigma^{2\alpha_I}}{\varrho_{H2U}^{\alpha_I}P_I^{\alpha_I}\Gamma(\alpha_I)\ln2}\int_{0}^{\infty}G^{1,2}_{2,2} \left(\zeta \middle| {^{1,1}_{1,0}}\right)\zeta^{\alpha_I-1}e^{-\frac{\beta_I\sigma^2 \zeta}{\varrho_{H2U}P_I}}\mathrm{d}\zeta\nonumber\\
            \overset{(c)}{=}&\frac{1}{\Gamma(\alpha_I)\ln 2}G^{1,3}_{3,2} \left(\frac{P_{I}\varrho_{H2U}}{\beta_I\sigma^2} \middle| {^{1-\alpha_I, 1, 1}_{1,\ \ 0}}\right),
        \end{align}
        where $(a)$ is due to variable substitution $\zeta = \varrho_{H2U}P_Iz/\sigma^2$. $(b)$ and $(c)$ are obtained by using \cite[Eq. (8.4.6.5)]{prudnikov1986integrals} and \cite[Eq. (2.24.3.1)]{prudnikov1986integrals}, respectively.
    \end{IEEEproof}

   Considering the imperfect DOA estimation, we have
    \begin{align}
         \mathcal{A}_{\sigma^2, N} =& \sum_{i=0}^{N-1}\sum_{k=0}^{N-1}e^{-\frac{(i-k)^2\sigma^2}{2}},\label{eqn:A2}\\
        \mathcal{A}^{(4)}_{\sigma^2,N}=&N^2+4N\sum_{i=1}^{N-1}(M-i)e^{-\frac{i^2\sigma^2}{2}}\nonumber\\
         &+2\sum_{i=1}^{N-1}\sum_{j=1}^{N-1}(N-i)(N-j)\left(e^{-\frac{(i-j)^2\sigma^2}{2}}+e^{-\frac{(i+j)^2\sigma^2}{2}}\right),\label{eqn:A4}
    \end{align}
where $\sigma^2\in\{\sigma_u^2, \sigma_v^2,\sigma_z^2\}$, $N\in\{N_x,N_y,M\}$. Since $\sigma^2_{u}=\sigma^2_{h,u}+\sigma^2_{G,u}$ and $N_x$, $\sigma^2_{v}=\sigma^2_{h,v}+\sigma^2_{G,v}$ and $N_y$, $\sigma^2_{z}=\sigma^2_{G,z}$ and $M$ always occur together, we can further simplify the notation by defining $\mathcal{A}_{\sigma_u^2, N_x}$ as $\mathcal{A}_u$, $\mathcal{A}_{\sigma_v^2, N_y}$ as $\mathcal{A}_v$, and $\mathcal{A}_{\sigma^2_z, M}$ as $\mathcal{A}_z$, and the same simplification applies to Eq. \eqref{eqn:A4}.

\begin{figure*}[hb]
    \hrulefill
    \begin{align}
        &\alpha_{E}^{DOA}=\frac{\left(\kappa_h\kappa_G\mathcal{A}_u\mathcal{A}_v\mathcal{A}_z+\kappa_G\mathcal{A}_zN+\kappa_hMN+MN\right)^2}
        {2\kappa_h\kappa_G\mathcal{A}_u\mathcal{A}_vN\left(\kappa_G\mathcal{A}_u^{(4)}+M\mathcal{A}_z\left(\kappa_h+1\right)\right)+\kappa_h^2\kappa_G^2\left(\prod\limits_{t}^{t=u,v,z}\mathcal{A}_t^{(4)}-\prod\limits_{t}^{t=u,v,z}\mathcal{A}^2_t\right)+\kappa_G\left(2N^2\mathcal{A}_z^{(4)}-N^2\mathcal{A}_z^2\right)+\kappa_h^2M^2N^2},\label{eqn:alphaErrorE}\\
        &\beta_{E}^{DOA}=\frac{(\varrho_{H2U}P_E)^{-1}\left(\kappa_h\kappa_G\mathcal{A}_u\mathcal{A}_v\mathcal{A}_z+\kappa_G\mathcal{A}_zN+\kappa_hMN+MN\right)(1+\kappa_h)(1+\kappa_G)}
        {2\kappa_h\kappa_G\mathcal{A}_u\mathcal{A}_vN\left(\kappa_G\mathcal{A}_u^{(4)}+M\mathcal{A}_z\left(\kappa_h+1\right)\right)+\kappa_h^2\kappa_G^2\left(\prod\limits_{t}^{t=u,v,z}\mathcal{A}_t^{(4)}-\prod\limits_{t}^{t=u,v,z}\mathcal{A}^2_t\right)+\kappa_G\left(2N^2\mathcal{A}_z^{(4)}-N^2\mathcal{A}_z^2\right)+\kappa_h^2M^2N^2}.\label{eqn:betaErrorE}
    \end{align}
\end{figure*}

\begin{Lemma}\label{lemma:lemma5}
The $E_{k}^{DOA}$ obeys the Gamma distribution with the shape parameter $\alpha_{E}^{DOA}$ and rate parameter $\beta_{E}^{DOA}$, where $\alpha_{E}^{DOA}$ and $\beta_{E}^{DOA}$ are given in Eq. \eqref{eqn:alphaErrorE} and Eq. \eqref{eqn:betaErrorE} at the bottom of this page.
\end{Lemma}
\begin{IEEEproof}
    Please refer to Appendix \ref{app:D} for detailed proof.
\end{IEEEproof}

\section{Numerical Result}\label{sec:sec4}
In this section, numerical results are presented to demonstrate the performance of the proposed RISS assisted WPCN system. In the simulations, the distance between the HAP and the RISS is fixed at 12 meters, while the distance between the RISS and the user is set to 3 meters. The signal attenuation is set to 30 dB at a reference distance of 1 meter, and the path loss exponent is set to 2.2 for both the HAP-to-RISS and RISS-to-user channels. Unless otherwise stated, the following setup is utilized: $M = 4$, $N = 100$ and the transmit power $P_E$ is set to 1 W.

\subsection{On the Impact of Rician Factor}
\begin{figure}
    \centering
    \includegraphics[width=0.8\linewidth]{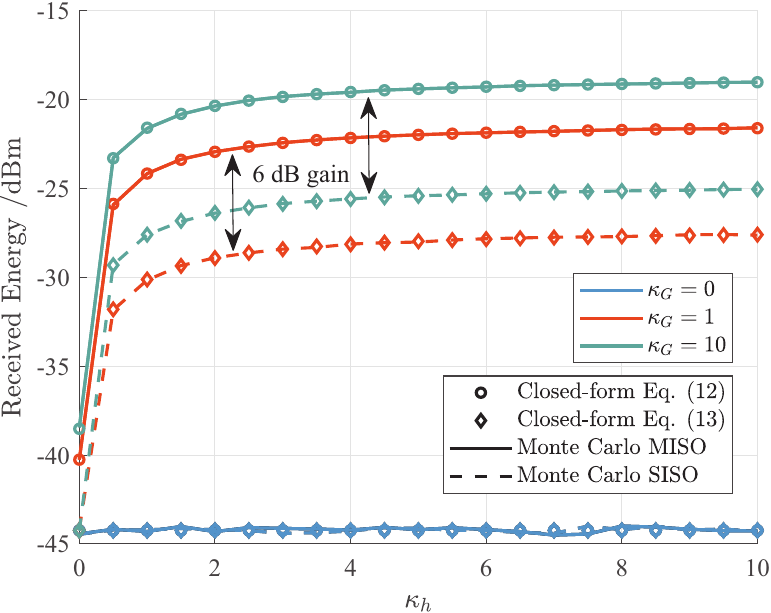}
    \setlength{\abovecaptionskip}{0pt}
    \setlength{\belowcaptionskip}{0pt} 
    \caption{The downlink performance of the proposed RISS assisted WPCN system exhibits variations with respect to the Rician factors $\kappa_h$ and $\kappa_G$. We consider a scenario where $M=4$ and $N=100$ to achieve a 6 dB improvement in MISO performance compared to SISO.}
    \label{fig:exp1}
\end{figure}
\begin{figure}
    \centering
    \includegraphics[width=0.8\linewidth]{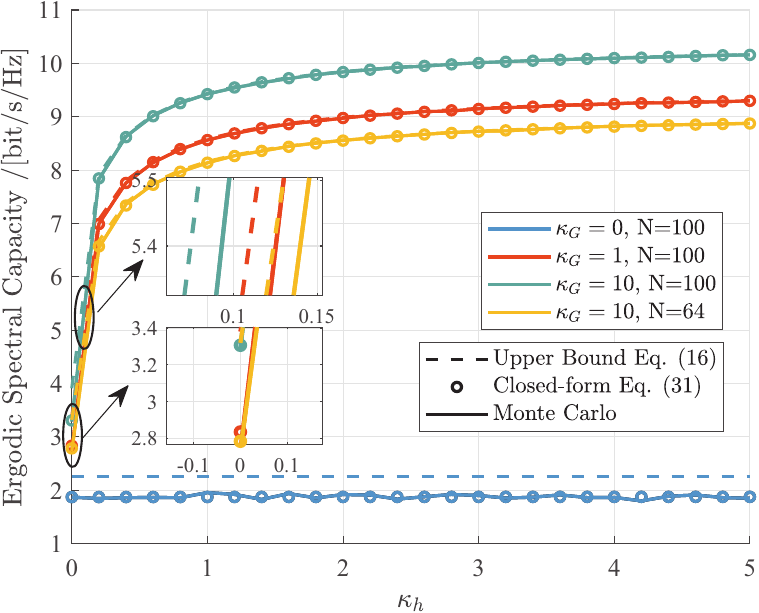}
    \setlength{\abovecaptionskip}{0pt}
    \setlength{\belowcaptionskip}{0pt} 
    \caption{The uplink ergodic spectral capacity performance of the proposed RISS assisted WPCN system varies with the Rician factors $\kappa_h$ and $\kappa_G$.}
    \label{fig:exp1_2}
\end{figure}

We first illustrate the performance discrepancy between SISO and MISO WET systems. Fig. \ref{fig:exp1} demonstrate that in the absence of LoS channel of $\mathbf{G}$ (i.e., $\kappa_G=0$), no discernible difference is observed between SISO and MISO, and the performance of both SISO and MISO has been kept at a low level with the increase of $\kappa_h$. However, in the scenario $\kappa_G\neq 0$, MISO system exhibit an increase of 6 dB (i.e., $M=4$ times) in received energy compared to the SISO counterpart, as we discussed deeply in Remark \ref{remark:1} and Table \ref{table:condratio}. Moreover, our numerical results reveal that the proposed scheme does not necessitate an excessively high value of $\kappa_h$, as a value greater than 3 is sufficient. The results highlight the significance of selecting the placement of the RISS to ensure the existence of $\kappa_G$ (e.g., it can be considered to deploy RISS at elevated location or in close proximity to HAP to obtain multi-antenna gain), even if its value is small (e.g., $\kappa_G=1$), followed by the guarantee of $\kappa_h$ to obtain a better performance. 

Fig. \ref{fig:exp1_2} displays the ergodic spectral capacity and the upper bound given in Eq. \eqref{eqn:upperbound_cc} and the closed-form given in Eq. \eqref{eqn:closedformcc2}, where the transmit power of single antenna user is set to $P_I=1$ mW, and noise power is set to $\sigma^2=-80$ dBm. The results show that when the channel conditions are ideal (e.g., $\kappa_G\neq0$ and $\kappa_h\geq 1$), the results of upper bound indicated by Eq. \eqref{eqn:upperbound_cc} and Monte Carlo simulations exhibit a high degree of consistency. However, as the channel conditions deteriorate (e.g., $\kappa_G=0$), the results diverge. And in all cases, the closed-form Eq. \eqref{eqn:closedformcc2} is highly consistent with the Monte Carlo results. Additionally, augmenting the number of passive elements in the RISS can significantly enhance the ergodic spectral efficiency.

\subsection{Performance with DOA Errors}
\begin{figure}
    \centering
    \includegraphics[width=0.8\linewidth]{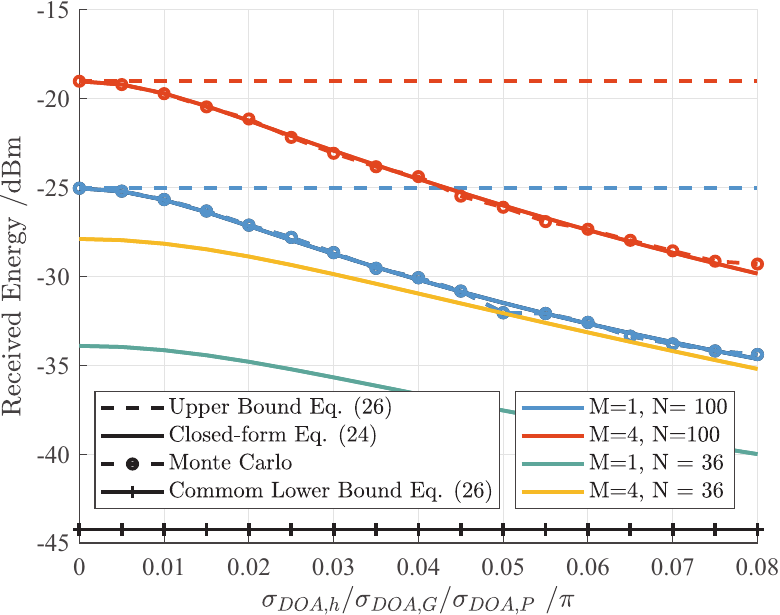}
    \setlength{\abovecaptionskip}{0pt}
    \setlength{\belowcaptionskip}{0pt} 
    \caption{The performance of the proposed scheme is evaluated in the presence of DOA errors. Both the MISO system (represented by the red line) and the SISO system (represented by the blue line) exhibit a common lower bound (represented by the green line) but different upper bounds in terms of received energy.}
    \label{fig:exp2}
\end{figure}

Fig. \ref{fig:exp2} demonstrates the capability of the closed-form expression, proposed in Eq. \eqref{eqn:err_final_closeform}, to accurately depict the changing trend of the mean value of energy reception with DOA estimation errors. As the variance of DOA errors increases gradually, the average received energy decreases. Specifically, both MISO and SISO systems share a common lower bound, and the performance benefit of MISO compared to SISO system decreases as the $\sigma_{DOA,P}$ increases, highlighting that the performance benefit between MISO and SISO can be ensured through the accurate acquisition of the $\varpi_G$ in HAP.

It is noteworthy that the robustness of the proposed scheme for DOA estimation errors decreases as the number of passive elements increases, since the beam will narrow as such an increase. Nevertheless, scale increase of RISS allows for greater space deployment of active elements, which in turn enhance the accuracy of DOA estimation.

\subsection{RISS assisted Versus full CSI}
\begin{figure}
    \centering
    \includegraphics[width=0.8\linewidth]{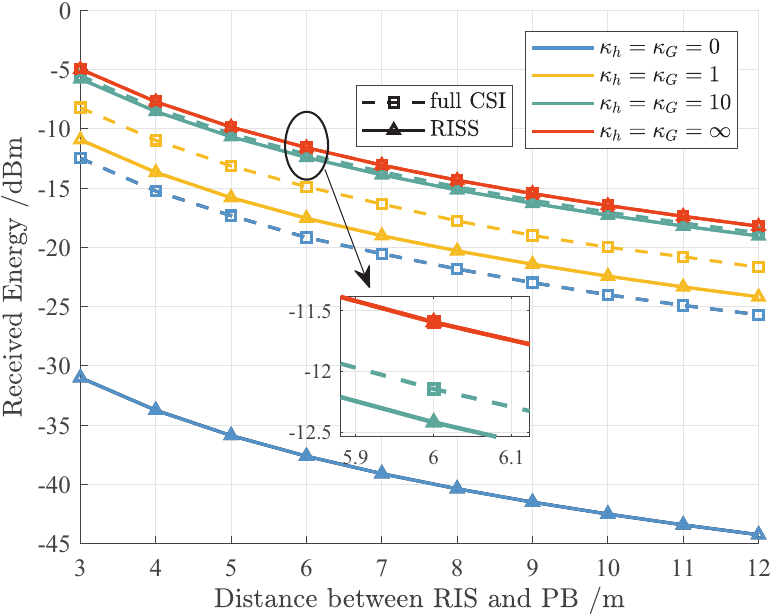}
    \setlength{\abovecaptionskip}{0pt}
    \setlength{\belowcaptionskip}{0pt} 
    \caption{The performance of the proposed scheme and the full CSI scheme varies with the distance between the RISS and HAP. We consider $N=4$ and $M=100$, and explore different values of $\kappa_G$ and $\kappa_h$ (0, 1, 10, or $\infty$) to examine their impact on the performance of both schemes.}
    \label{fig:exp3}
\end{figure}

\begin{figure}
    \centering
    \includegraphics[width=0.8\linewidth]{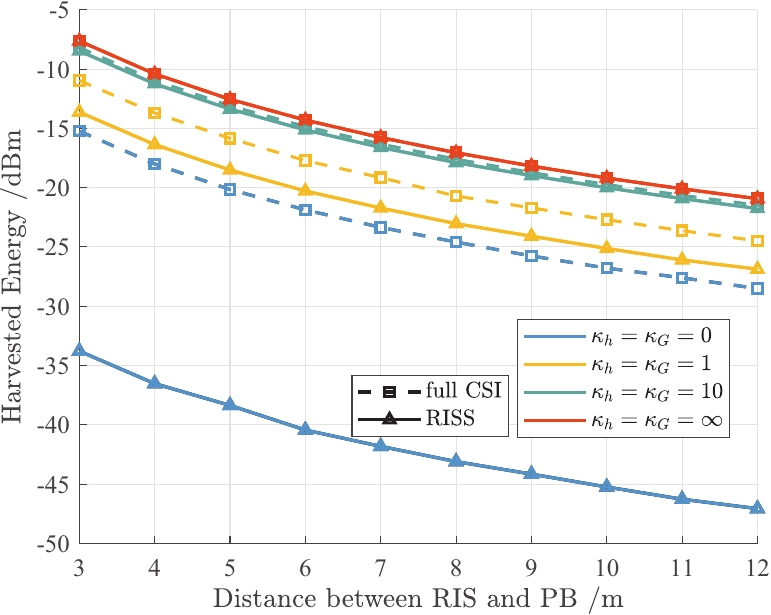}
    \setlength{\abovecaptionskip}{0pt}
    \setlength{\belowcaptionskip}{0pt} 
    \caption{The performance of the proposed scheme and the full CSI counterpart, both employing a non-linear energy harvesting model, where $M_e$ denote saturation harvested power, $a$ and $b$ are constants related to the detailed circuit specifications.}
    \label{fig:nonlinearenergymodel}
\end{figure}

\begin{figure}
    \centering
    \includegraphics[width=0.8\linewidth]{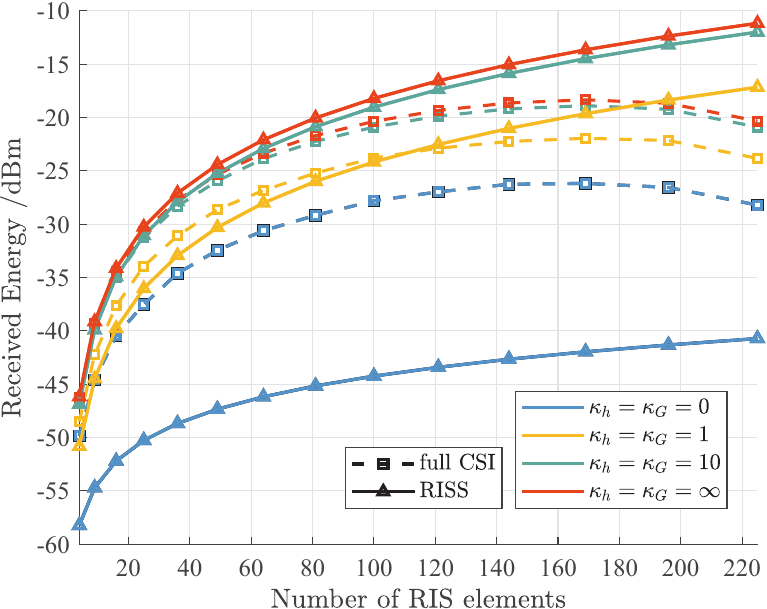}
    \setlength{\abovecaptionskip}{0pt}
    \setlength{\belowcaptionskip}{0pt} 
    \caption{The performance of the proposed scheme and the full CSI scheme is evaluated taking into account the pilot overhead. We consider a linear pilot overhead model and investigate the impact of different values of $\kappa_G$ and $\kappa_h$ (0, 1, 10, or $\infty$) on the performance of both schemes.}
    \label{fig:exp4}
\end{figure}
Fig. \ref{fig:exp3} illustrate the performance difference between our proposed scheme and full CSI scheme\cite{fullCSI_wqq}. It is important to highlight that the full CSI scheme operates under the assumption of perfect CSI. This scheme involves iteratively optimizing the transmission precoder at the HAP and the reflection phase shift of the RIS until convergence, all with the aim of maximizing the energy received by the receiver. When we have Rayleigh fading in both HAP-to-RISS and RISS-to-user channels (i.e., $\kappa_h=\kappa_G=0$), DOA estimation is invalid, since a significant performance gap between the proposed scheme and the full CSI scheme is observed in Fig. \ref{fig:exp3}. Nevertheless, if the LoS path is present, this performance gap quickly diminishes. For example, when $\kappa_h=\kappa_G=10$, the gap totally eliminates. This demonstrates that when the LoS path is strong, full CSI is not indispensable.

Generally, the received energy means the incident energy at the receiver, and the harvested energy denotes the energy through an energy harvester. To better illustrate the impact of the non-linear energy harvesting model \cite{nonlinearHarvesting} in our system, we have conducted additional counterpart experiments. Specifically, Fig. \ref{fig:nonlinearenergymodel} shows the harvested energy with parameters $M_e=0.02337, a= 132.8, b = 0.01181$ \cite{nonlinear} and others are the same parameter as Fig. \ref{fig:exp3}. While the values vary, note that the trend in the received energy and the harvested energy remains consistent even after introducing the non-linear energy harvesting model. This consistency arises from the fact that the distribution of energy reception falls within the linear range of the energy receiver, which allows us to focus exclusively on the performance of received energy in subsequent content.

Considering the pilot overhead, we know that as the number of RIS passive elements increases, pilot overhead will consume a substantial amount of channel correlation time, resulting in a significant reduction in available energy transmission time\cite{pilotoverhead1,pilotoverhead2}. To model this phenomenon, we have adopted a simple linear approach wherein the pilot overhead $T_p$ proportionally with the number of RIS passive elements, such that $T_p=N+1$ \cite{pilot_linear}. Specifically, we have made the assumption that $\mathbf{h}_k$ and $\mathbf{G}$ remain constant over a channel coherence block of length $T_c=256$, with $T_p<T_c$. Therefore, the remaining time available for energy transfer is given by $T_c-T_p=T_c-N-1$.

Fig. \ref{fig:exp4} further illustrates previous problem.
Firstly, the performance of both the proposed scheme and its full CSI counterpart augments with an increase in the number of passive elements of the RIS. However, as mentioned previously, the pilot overhead poses a detrimental impact on the transmission performance, especially when $N>100$. Besides, our proposed RISS scheme still maintain the trend of growth with $N$. Therefore, it can be deduced that our proposed scheme is more suitable for large-scale RIS.

Moreover, a fascinating conclusion is that we cannot increase the received energy and SNR without limitation by increasing the number of RIS passive elements, since the overhead of the pilot is also become intolerable. This demonstrates that there is an optimal RIS scale for the full CSI scheme to maximize the received energy and SNR.

\subsection{Outage Probability Performance}
\begin{figure}
    \centering
    \includegraphics[width=0.8\linewidth]{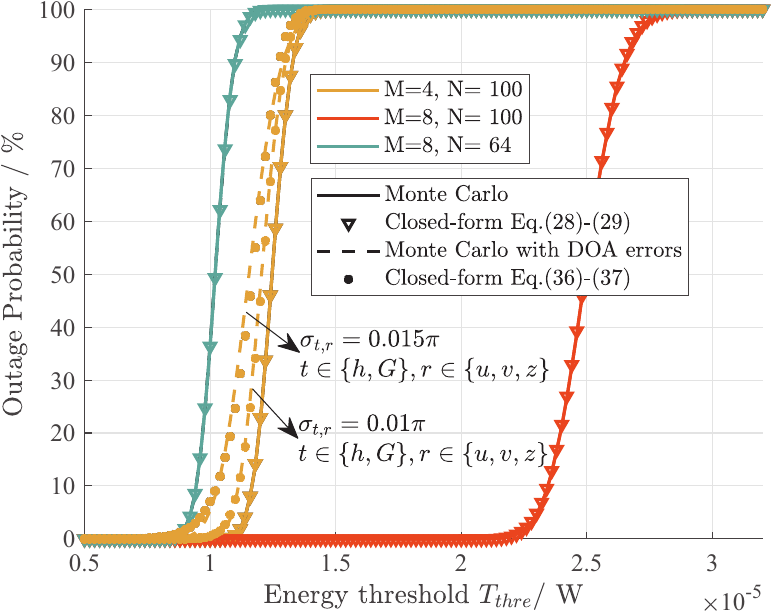}
    \setlength{\abovecaptionskip}{0pt}
    \setlength{\belowcaptionskip}{0pt} 
    \caption{The outage probability of proposed scheme varies with $M$, $N$ and required energy threshold. The transmit power is set to 1 W, while $\kappa_h$ and $\kappa_G$ are both set to 10.}
    \label{fig:exp5}

\end{figure}

\begin{figure}
    \centering
    \includegraphics[width=0.8\linewidth]{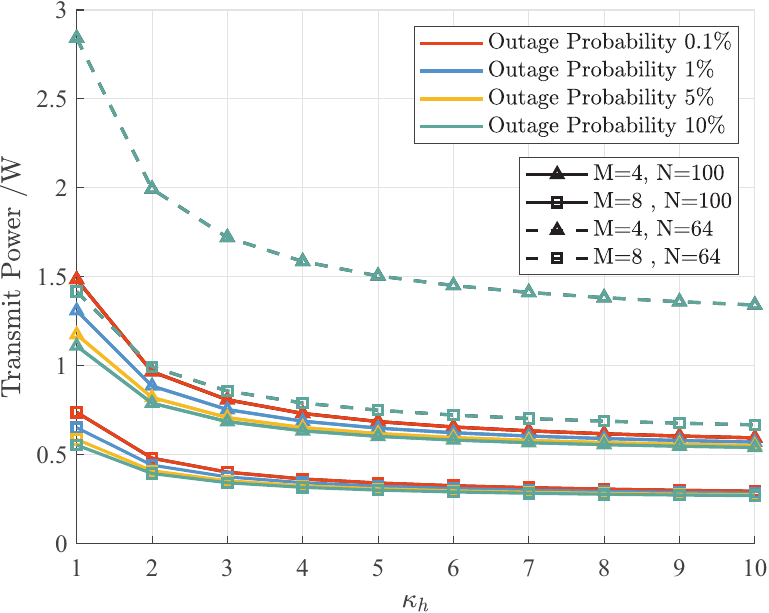}
    \setlength{\abovecaptionskip}{0pt}
    \setlength{\belowcaptionskip}{0pt} 
    \caption{The transmit power of proposed scheme varies with $M$, $N$ and $\kappa_h$, while keeping $\kappa_G$ fixed at 10 and the required energy threshold set to -22 dBm.}
    \label{fig:exp5_2}
\end{figure}
Fig. \ref{fig:exp5} depicts the variation in outage probability for different values of $M$ and $N$, where both $\kappa_h$ and $\kappa_G$ are set to 10. The correctness of the closed-form analysis obtained in Lemma \ref{lemma:lemma3} can be deduced from its close consistency with the Monte Carlo simulations on all test values. Note that the increase of $M$ and $N$ has significant impact on the performance of energy reception, while the range of received energy becomes wider as $N$ increases, which also means that the variance of received energy grows. Furthermore, we also illustrate the outage probability in the presence of DOA estimation errors. Observe from Fig. \ref{fig:exp5} that the closed-form expression in Lemma \ref{lemma:lemma5} effectively characterizes the actual energy reception distribution when the errors are small (e.g., $\sigma_{t,r}=0.01\pi,t\in\{h,G\},r\in\{u,v,z\}$). However, as the DOA estimation errors increase (e.g., $\sigma_{t,r}=0.015\pi,t\in\{h,G\},r\in\{u,v,z\}$), discrepancies emerge between the closed-form expression and the actual energy reception distribution. Within a reasonable range of errors in practical systems, these discrepancies are neglected.

Fig. \ref{fig:exp5_2} displays the transmit power required under various outage probability demand with accurate DOA estimation and different $\kappa_h$, where $\kappa_G=10$. The numerical results shows that the increase of $M$, $N$ and $\kappa_{h}$ significant decreases the transmit power. Furthermore, in the case of fixed $M$ and $N$, increasing $\kappa_h$ often leads to the convergence of the transmit power to similar levels. This observation indicates a reduction in the variance of received energy. Additionally, it is important to note that meeting the outage requirement from 1\% to 0.1\% for the transmitter entails much higher energy requirements compared to that of meeting the requirement from 10\% to 5\%, as the CDF shows a gradual rising trend towards the tail. This suggests that setting excessively strict outage probability conditions is not feasible, as it would result in unacceptable energy consumption.
\section{Conclusion}\label{sec:sec5}

This paper introduces a RISS assisted WPCN system designed to appreciably decrease pilot overhead and decoupling from the HAP through the concurrent execution of reflection and DOA estimation tasks. The paper initially presents the system and protocol designs, succeeded by the derivation of the average WET performance and the determination of the stringent upper bound of WIT. The DOA estimation errors are subsequently scrutinized, and a closed-form expression is derived for performance evaluation. Additionally, the statistical distribution of user-received energy is ascertained through the moment-matching technique, and the optimal transmit power for the HAP is derived. To substantiate the proposed scheme, a series of comprehensive experiments is performed, which verify its superiority over the full CSI counterpart under favorable channel conditions with a substantial count of RIS passive elements. The proposed system, by virtue of its reduced implementation complexity and enhanced efficiency, holds practical significance for the optimization of WPCN systems.

{\appendices
\section{Proof of Eq. \eqref{eqn:not_cons_h}-Eq. \eqref{eqn:not_cons_G}}\label{app:A}
To prove this conclusion, we first rewrite Eq. \eqref{eqn:vectoralpha} as
\begin{align}
    &\boldsymbol{\alpha}(u, v) = \boldsymbol{\alpha}_x(u)\otimes \boldsymbol{\alpha}_y(v),
\end{align}
where $\boldsymbol{\alpha}_x(u)=\frac{1}{\sqrt{N_x}}\left[1,\cdots,e^{(N_x-1)\mathbbm{i}u}\right]^\mathrm{T}$, $\boldsymbol{\alpha}_y(v)=\frac{1}{\sqrt{N_y}}\left[1,\cdots,e^{(N_y-1)\mathbbm{i}v}\right]^\mathrm{T}$, $u = \pi\cos(\varphi)$ and $v=\pi\sin(\varphi)\sin(\vartheta)$. Thus, Eq. \eqref{eqn:hG_cons} can be expressed as
\begin{align}
    &\bar{\mathbf{h}}^\mathrm{H}\mathrm{diag}\{\theta_{1,{h}},\theta_{2,{h}},\cdots,\theta_{N,{h}}\}\nonumber\\
    \overset{(a)}{=}&\sqrt{N}(\boldsymbol{\alpha}_{x}^\mathrm{H}(u_h)\otimes\boldsymbol{\alpha}_{y}^\mathrm{H}(v_h))(\sqrt{N_x}\boldsymbol{\alpha}_{x}(u^e_h)\otimes\sqrt{N_y}\boldsymbol{\alpha}_{y}(v^e_h))\nonumber\\
    \overset{(b)}{=}&N\left(\boldsymbol{\alpha}_{x}^\mathrm{H}(u_h)\boldsymbol{\alpha}_{x}(u^e_h)\right)\otimes\left(\boldsymbol{\alpha}_{y}^\mathrm{H}(v_h)\boldsymbol{\alpha}_{y}(v^e_h)\right)\nonumber\\
    \overset{(c)}{=}&\sqrt{N}\boldsymbol{\alpha}^\mathrm{H}(\xi_{h,u},\xi_{h,v}),
\end{align}
where $(a)$ and $(b)$ comes from $(\mathbf{A}\otimes \mathbf{B})^\mathrm{H}=\mathbf{A}^\mathrm{H}\otimes \mathbf{B}^\mathrm{H}$ and $(\mathbf{A}\otimes\mathbf{B})(\mathbf{C}\otimes\mathbf{D})=\mathbf{(AC)}\otimes\mathbf{(BD)}$, and $(c)$ is due to the fact $u_h-u_h^e=\xi_{h,u}$ and $v_h-v_h^e=\xi_{h,v}$. Similarly, this holds true for
\begin{align}
    &\mathrm{diag}\{\theta_{1,{G}},\cdots,\theta_{N,{G}}\}\bar{\mathbf{G}}\mathbf{w}\nonumber\\
    =&\left(\sqrt{N_x}\boldsymbol{\alpha}_{x}(-u^e_G)\otimes\sqrt{N_y}\boldsymbol{\alpha}_{y}(-v^e_G)\right)\nonumber\\
    &\qquad\qquad\times
    \left(\sqrt{MN}\boldsymbol{\alpha}_{x}(u_G)\otimes\boldsymbol{\alpha}_{y}(v_G)\boldsymbol{\beta}^\mathrm{H}(z_{G})\right)
    \frac{\sqrt{P_E}\boldsymbol{\beta}(z_{G})}{||\boldsymbol{\beta}(z_{G})||}\nonumber\\
    =& \sqrt{MNP_E}\boldsymbol{\alpha}(\xi_{G,u},\xi_{G,v})\frac{1}{\sqrt{M}}\mathbf{1}^\mathrm{T}_{M\times 1}\boldsymbol{\beta}(\xi_{G,z}).
\end{align}
Then the proof is then complete.
\section{Proof of Lemma \ref{lemma:lemma1}}\label{app:B}
Expand Eq. \eqref{eqn:error_cons} we have
\begin{align}
    &\left|\sqrt{N}\mathbf{1}_{N\times 1}^\mathrm{T}\boldsymbol{\alpha}(\xi_{h,u}+\xi_{G,u},\xi_{h,v}+\xi_{G,v})\mathbf{1}^\mathrm{T}_{M\times 1}\boldsymbol{\beta}(\xi_{G,z})\right|^2\nonumber\\
    =&\left|\sum_{n_x=0}^{N_x-1}e^{\mathbbm{i}n_x(\xi_{h,u}+\xi_{G,u})}\sum_{n_y=0}^{N_y-1}e^{\mathbbm{i}n_y(\xi_{h,v}+\xi_{G,v})}\frac{1}{\sqrt{M}}\sum_{m=0}^{M-1}e^{\mathbbm{i}m\xi_{G,z}}\right|^2\nonumber\\
    =&\underbrace{\left|\sum_{n_x=0}^{N_x-1}e^{\mathbbm{i}n_x(\xi_{h,u}+\xi_{G,u})}\right|^2}_{\text{Part 1}}\underbrace{\left|\sum_{n_y=0}^{N_y-1}e^{\mathbbm{i}n_y(\xi_{h,v}+\xi_{G,v})}\right|^2}_{\text{Part 2}}\frac{1}{M}\underbrace{\left|\sum_{m=0}^{M-1}e^{\mathbbm{i}m\xi_{G,z}}\right|^2}_{\text{Part 3}}.
\end{align}

We can observe that Part 1-3 have the same expression form. And for Part 3, we have
\begin{align}
    \mathbb{E}\left\{\left|\sum_{m=0}^{M-1}e^{\mathbbm{i}m\xi_{G,z}}\right|^2\right\}
    \overset{(a)}{=}&\mathbb{E}\left\{\sum_{i=0}^{M-1}\sum_{k=0}^{M-1}\cos\left((i-k)\xi_{G,z}\right)\right\}\nonumber\\
    \overset{(b)}{=}&\sum_{i=0}^{M-1}\sum_{k=0}^{M-1}\sum_{n=0}^\infty\frac{(-1)^{n}}{(2n)!}\mathbb{E}\left\{\left((i-k)\xi_{G,z}\right)^{2n}\right\}.\label{eqn:taylorseries}
\end{align}
Where $(a)$ is due to Euler's formula and prosthaphaeresis, $(b)$ comes from the Taylor series. We make $\Psi=\left((i-k)\xi_{G,z}\right)^{2n}$, and since $\xi_{G,z}\sim\mathcal{N}(0,\sigma^2_{G,z})$, we have the probability density function (PDF) of $\Psi$ is
\begin{align}
    f_{\Psi}(\psi)=f_{\xi_{G,z}}(\psi^{\frac{1}{2n}})\frac{1}{n}\psi^{\frac{1}{2n}-1}=\frac{\psi^{\frac{1}{2n}-1}}{\sqrt{2\pi} n\sigma_{G,z}}e^{-\frac{\psi^{\frac{1}{n}}}{2\sigma_{G,z}^2}},
\end{align}
where $f_{\Psi}(\cdot)$ and $f_{\xi_{G,z}}(\cdot)$ represent the PDF of variable $\Psi$ and $\xi_{G,z}$, respectively. Thus we hare
\begin{align}
    &\mathbb{E}\left\{\left((i-k)\xi_{G,z}\right)^{2n}\right\}\nonumber\\
    =&\int_0^\infty \psi f_{\Psi}(\psi)\text{d} \psi=\int_0^\infty\frac{\psi^{\frac{1}{2n}}}{\sqrt{2\pi} n(i-k)\sigma_{G,z}}e^{-\frac{\psi^{\frac{1}{n}}}{2(i-k)^2\sigma_{G,z}^2}}\text{d}\psi\nonumber\\
    =&\frac{(2n)!}{n!2^n}\left((i-k)\sigma_{G,z}\right)^{2n}.\label{eqn:expeofvarXI}
\end{align}
Take Eq. \eqref{eqn:expeofvarXI} into Eq. \eqref{eqn:taylorseries}, yielding
\begin{align}
    &\sum_{i=0}^{M-1}\sum_{k=0}^{M-1}\sum_{n=0}^\infty\frac{(-1)^{n}}{(2n)!}\mathbb{E}\left\{\left((i-k)\xi_{G,z}\right)^{2n}\right\}\nonumber\\
    =&\sum_{i=0}^{M-1}\sum_{k=0}^{M-1}\sum_{n=0}^\infty\frac{(-1)^{n}}{(2n)!}\frac{(2n)!}{n!2^n}\left((i-k)\sigma_{G,z}\right)^{2n}\nonumber\\
    =&\sum_{i=0}^{M-1}\sum_{k=0}^{M-1}e^{-\frac{(i-k)^2\sigma_{G,z}^2}{2}}.\label{eqn:sumsumexp}
\end{align}
Similarly, this holds true for both Part 1 and Part 2, thereby completing the proof.

\section{Proof of Lemma \ref{lemma:lemma3}}\label{app:C}
Denote by ${\omega}_1=\sqrt{\kappa_h\kappa_G}N\sqrt{M}$, ${\omega}_2=\sqrt{\kappa_GM}\hat{\mathbf{h}}^\S\mathbf{1}_{N\times1}\sim\mathcal{CN}(0, \kappa_GMN)$, ${\omega}_3=\sqrt{\kappa_h}\mathbf{1}_{N\times1}^\mathrm{T}\hat{\mathbf{G}}^\S\sim\mathcal{CN}(0,\kappa_hN)$, $\omega_4 = \sum_{i=1}^N[\hat{\mathbf{h}}^\S]_i[\hat{\mathbf{G}}^\S]_i$, Eq. \eqref{eqn:insE} can be rewritten as
\begin{align}
    \frac{\varrho_{H2U}P_E}{(1+k_h)(1+k_G)}\left|\omega_1+\omega_2+\omega_3+\omega_4\right|^2.\label{eqn:rewrittenB}
\end{align}
Since the mean value of Eq. \eqref{eqn:rewrittenB} are given by Eq. \eqref{eqn:cons_sta}, the variance can be expressed as
\begin{align}
    &\frac{\varrho_{H2U}^2P_E^2}{(1+k_h)^2(1+k_G)^2}\mathbb{D}\left\{\sum_{i=1}^4\omega_i^\mathrm{H}\omega_i+2\sum_{i=1}^3\sum_{j=i+1}^4\Re\{\omega_i^\mathrm{H}\omega_j\}\right\}\nonumber\\
    =&\frac{\varrho_{H2U}^2P_E^2}{(1+k_h)^2(1+k_G)^2}\left(\sum_{i=1}^4\mathbb{D}\left\{\omega_i^\mathrm{H}\omega_i\right\}+4\sum_{i=1}^3\sum_{j=i+1}^4\mathbb{D}\left\{\Re\{\omega_i^\mathrm{H}\omega_j\}\right\}\right. \nonumber\\
    &\left.+2\left(\text{Cov}_{2,2;4,4}+\text{Cov}_{3,3;4,4}+\text{Cov}_{1,2;3,4}+\text{Cov}_{1,3;2,4}\right)
    \right).\label{eqn:varexpand}
\end{align}
Where $\mathbb{D}\left\{\omega_1^\mathrm{H}\omega_1\right\}=0$ and
\begin{align}
    \mathbb{D}\left\{\omega_2^\mathrm{H}\omega_2\right\}=\mathbb{D}\left\{\left|\omega_2\right|^2\right\}=\kappa_G^2M^2N^2,\label{eqn:accw2w2}
\end{align}
where $\left|\omega_2\right|^2\sim\mathcal{X}^2_{2,\kappa_GMN/2}$. Similarly, we have
\begin{align}
    \mathbb{D}\left\{\omega_3^\mathrm{H}\omega_3\right\}=\mathbb{D}\left\{\left|\omega_3\right|^2\right\}=\kappa_h^2N^2.
\end{align}
And for $\mathbb{D}\left\{\omega_4^\mathrm{H}\omega_4\right\}$ we have
\begin{align}
    \mathbb{D}\left\{\omega_4^\mathrm{H}\omega_4\right\}=&\mathbb{D}\left\{\sum_{i=1}^N[\hat{\mathbf{h}}^{\S, \mathrm{H}}]_i[\hat{\mathbf{G}}^{\S,\mathrm{H}}]_i\sum_{j=1}^N[\hat{\mathbf{h}}^\S]_j[\hat{\mathbf{G}}^\S]_j\right\}\nonumber\\
    =&
    \begin{cases}
        \mathbb{D}\left\{\sum_{i=1}^N\left|[\hat{\mathbf{h}}^\S]_i\right|^2\left|[\hat{\mathbf{G}}^\S]_i\right|^2\right\}, i=j\nonumber\\
        \mathbb{D}\left\{\sum^N_{i,j=1}[\hat{\mathbf{h}}^{\S,\mathrm{H}}]_i[\hat{\mathbf{h}}]_j[\hat{\mathbf{G}}^{\S,\mathrm{H}}]_i[\hat{\mathbf{G}}^\S]_j\right\}, i\neq j\nonumber\\
    \end{cases}\\
    \overset{(a)}{=}&3N+N^2-N=N^2+2N,\label{eqn:appdixc_varexpan}
\end{align}
where $(a)$ comes from $\mathbb{D}\{XY\}=\mathbb{D}\{X\}\mathbb{D}\{Y\}+\mathbb{D}\{X\}\mathbb{E}^2\{Y\}+\mathbb{D}\{Y\}\mathbb{E}^2\{X\}$.
And since
\begin{align}
    4\mathbb{D}\left\{\Re\{\omega_1^\mathrm{H}\omega_2\}\right\}=&4\kappa_G^2\kappa_hN^2M^2\mathbb{D}\left\{\Re\{\hat{\mathbf{h}}^\mathrm{H}\mathbf{1}_{N\times1}\}\right\}\nonumber\\
    =&2\kappa_h\kappa_G^2M^2N^3,
\end{align}
we can obtain
\begin{align}
    &4\mathbb{D}\left\{\Re\{\omega_1^\mathrm{H}\omega_3\}\right\}=2\kappa_h^2\kappa_GMN^3,\nonumber\\
    &4\mathbb{D}\left\{\Re\{\omega_1^\mathrm{H}\omega_4\}\right\}=2\kappa_h\kappa_GMN^3,\nonumber\\
    &4\mathbb{D}\left\{\Re\{\omega_3^\mathrm{H}\omega_4\}\right\}=2\kappa_h(N^2+N),\nonumber\\
    &4\mathbb{D}\left\{\Re\{\omega_2^\mathrm{H}\omega_3\}\right\}=2\kappa_G\kappa_hMN^2,
\end{align}
And $4\mathbb{D}\left\{\Re\{\omega_2^\mathrm{H}\omega_4\}\right\}$ can be calculated similarly with Eq. \eqref{eqn:appdixc_varexpan} as
\begin{align}
    4\mathbb{D}\left\{\Re\{\omega_2^\mathrm{H}\omega_4\}\right\}=&2\kappa_GM\times
    \begin{cases}
        \mathbb{D}\left\{\sum_{i=1}^N\left|[\hat{\mathbf{h}}^\S]_i\right|^2[\hat{\mathbf{G}}^{\S}]_i\right\},i=j\\
        \mathbb{D}\left\{\sum^N_{\substack{i,j=1\\i\neq j}}[\hat{\mathbf{h}}^\S]_i[\hat{\mathbf{G}}^{\S}]_i[\hat{\mathbf{h}}^\S]_j\right\},i\neq j
    \end{cases}\nonumber\\
    =&2\kappa_GM(N^2+N).
\end{align}
The covariance terms can be expanded as
\begin{align}
    \text{Cov}_{2,2;4,4}=&\mathbb{E}\left\{\omega_2^\mathrm{H}\omega_2\omega_4^\mathrm{H}\omega_4\right\}-\mathbb{E}\left\{\omega_2^\mathrm{H}\omega_2\right\}\mathbb{E}\left\{\omega_4^\mathrm{H}\omega_4\right\}\nonumber\\
    =&\mathbb{D}\left\{\omega_2^\mathrm{H}\omega_4\right\}+\mathbb{E}^2\left\{\omega_2^\mathrm{H}\omega_4\right\}-\kappa_GMN^2=\kappa_GMN,
\end{align}
and similarly we have
\begin{align}
    &\text{Cov}_{3,3;4,4}=\kappa_hN,\nonumber\\
    &\text{Cov}_{1,2;3,4}=2\kappa_h\kappa_GMN^2,\nonumber\\
    &\text{Cov}_{1,3;2,4}=2\kappa_h\kappa_GMN^2,\label{eqn:appc_var_final}
\end{align}
Ultimately, we employ the moment-matching approach to equate the CDF of $E_{k}$ with a Gamma distribution, where the shape and rate parameters are determined as $\alpha_E=\frac{\mathbb{E}^2\left\{E_{k}\right\}}{\mathbb{D}\left\{E_{k}\right\}}$ and $\beta_E=\frac{\mathbb{E}\left\{E_{k}\right\}}{\mathbb{D}\left\{E_{k}\right\}}$, respectively. By utilizing Eq. \eqref{eqn:cons_sta}  and Eq. \eqref{eqn:appc_var_final} in the aforementioned expressions, we derive the outcome presented in Lemma \ref{lemma:lemma3}.

\section{Proof of Lemma \ref{lemma:lemma5}}\label{app:D}
We continue to assume that the DOA estimation error follows a Gaussian distribution in the spatial phase differences $(u,v,z)$, Eq. \eqref{eqn:varexpand} can be formulated as
\begin{align}
    &\frac{\varrho_{H2U}^2P_E^2}{(1+k_h)^2(1+k_G)^2}\mathbb{D}\left\{\sum_{i=1}^4\omega_i^\mathrm{H}\omega_i+2\sum_{i=1}^3\sum_{j=i+1}^4\Re\{\omega_i^\mathrm{H}\omega_j\}\right\}\nonumber\\
    &\overset{(a)}{\approx}\frac{\varrho_{H2U}^2P_E^2}{(1+k_h)^2(1+k_G)^2}\left(\sum_{i=1}^3\mathbb{D}\left\{\omega_i^\mathrm{H}\omega_i\right\}\right.\nonumber\\
    &\qquad\qquad\qquad\qquad\qquad\qquad\left.+4\sum_{j=2}^4\mathbb{D}\left\{\Re\{\omega_1^\mathrm{H}\omega_j\}\right\}
    \right).\label{eqn:varexpand2}
\end{align}
where $(a)$ is approximated because we neglect some smaller contributing components according to Eq. \eqref{eqn:accw2w2}-\eqref{eqn:appc_var_final} when $\kappa_h\neq 0$ and $\kappa_G\neq 0$. Moreover, we have
\begin{align}
&\omega_1=N\sqrt{\kappa_h\kappa_G}\boldsymbol{\alpha}^\mathrm{H}(\xi_{h,u},\xi_{h,v})\boldsymbol{\alpha}(\xi_{G,u},\xi_{G,v})\mathbf{1}_{M\times 1}^\mathrm{T}\boldsymbol{\beta}(\xi_{G,z}),\nonumber\\ 
&\omega_2=\sqrt{N\kappa_G}\hat{\mathbf{h}}^{\S}\boldsymbol{\alpha}(\xi_{G,u},\xi_{G,v})\mathbf{1}_{M\times 1}^\mathrm{T}\boldsymbol{\beta}(\xi_{G,z}),\nonumber\\
&\omega_3=\sqrt{N\kappa_h}\boldsymbol{\alpha}^\mathrm{H}(\xi_{G,u},\xi_{G,v})\hat{\mathbf{G}}^{\S},\nonumber\\
&\omega_4=\hat{\mathbf{h}}^{\S}\hat{\mathbf{G}}^{\S}.
\end{align}
Since $\mathbb{E}\{\hat{\mathbf{h}}^{\S}\}=0$, we further have
\begin{align}
    \mathbb{D}\{\omega_1^\mathrm{H}\omega_2\}=&\mathbb{E}\left\{\left|\omega_1^\mathrm{H}\omega_2\right|^2\right\}-\mathbb{E}\left\{\omega_1^\mathrm{H}\omega_2\right\}^2=\mathbb{E}\left\{\left|\omega_1^\mathrm{H}\omega_2\right|^2\right\}\nonumber\\
    =&\kappa_h\kappa_G^2\mathbb{E}\left\{\left|\sqrt{N}\boldsymbol{\alpha}^\mathrm{H}(\xi_{h,u}+\xi_{G,u},\xi_{h,v}+\xi_{G,v})\right.\right.\nonumber\\
    &\left.\left.\qquad\times\mathbf{1}_{N\times1}\left(\mathbf{1}^\mathrm{T}_{M\times1}\boldsymbol{\beta}(\xi_{G,z})\right)^2\hat{\mathbf{h}}^{\S}\boldsymbol{\alpha}(\xi_{G,u},\xi_{G,v})\right|^2\right\}\nonumber\\
    =&\kappa_h\kappa_G^2\mathcal{A}_u\mathcal{A}_v\mathbb{E}\left\{\left|\sum_{m=0}^{M-1}e^{\mathbbm{i}m\xi_{G,z}}\right|^4\right\}N,\label{eqn:w1w2}
\end{align}
where $\mathcal{A}_u$, $\mathcal{A}_v$ are defined in Eq. \eqref{eqn:A2} and Eq. \eqref{eqn:A4}. And 
\begin{align}
    \mathcal{A}_{z}^{(4)}=&\mathbb{E}\left\{\left|\sum_{m=0}^{M-1}e^{\mathbbm{i}m\xi_{G,z}}\right|^4\right\}=\mathbb{E}\left\{\left(\sum_{i=0}^{M-1}\sum_{k=0}^{M-1}\cos\left((i-k)\xi_{G,z}\right)\right)^2\right\}\nonumber\\
    =&\mathbb{E}\left\{\left(M+2\sum_{m=1}^{M-1}(M-m)\cos(m\xi_{G,z})\right)^2\right\}\nonumber\\
    \overset{(a)}{=}&M^2+4M\sum_{m=1}^{M-1}(M-m)e^{-\frac{m^2\sigma^2_{G,z}}{2}}\nonumber\\
    &+2\sum_{m=1}^{M-1}\sum_{n=1}^{M-1}(M-m)(M-n)\left(e^{-\frac{(m-n)^2\sigma_{G,z}^2}{2}}+e^{-\frac{(m+n)^2\sigma_{G,z}^2}{2}}\right),
\end{align}
where $(a)$ comes from $\cos(a)\cos(b)=\frac{1}{2}\left(\cos(a-b)+\cos(a+b)\right)$ and Eq. \eqref{eqn:sumsumexp}. Then Eq. \eqref{eqn:w1w2} can be finally expressed as
\begin{align}
    \mathbb{D}\left\{\Re\left\{\omega_1^\mathrm{H}\omega_2\right\}\right\} = \frac{1}{2}\mathbb{E}\left\{\left|\omega_1^\mathrm{H}\omega_2\right|^2\right\}=\frac{\kappa_h\kappa_G^2}{2M^2}\mathcal{A}_u\mathcal{A}_v\mathcal{A}_{z}^{(4)}N. \label{eqn:finalw1w2}
\end{align}
And similarly we have 
\begin{align}
    &\mathbb{D}\left\{\Re\left\{\omega_1^\mathrm{H}\omega_3\right\}\right\}=\frac{\kappa_h^2\kappa_G}{2M}\mathcal{A}_u\mathcal{A}_v\mathcal{A}_zN,\nonumber\\
    &\mathbb{D}\left\{\Re\left\{\omega_1^\mathrm{H}\omega_4\right\}\right\}=\frac{\kappa_h\kappa_G}{2M}\mathcal{A}_u\mathcal{A}_v\mathcal{A}_zN,
\end{align}
as well as
\begin{align}
	\mathbb{D}\left\{\omega_1^\mathrm{H}\omega_1\right\}&=\mathbb{E}\left\{\left|\omega_1\right|^4\right\}-\mathbb{E}\left\{\left|\omega_1\right|^2\right\}^2\nonumber\\
    &=\frac{\kappa_h^2\kappa_G^2}{M^2}\left(\mathcal{A}_{u}^{(4)}\mathcal{A}_{v}^{(4)}\mathcal{A}_{z}^{(4)}-\mathcal{A}_{u}^2\mathcal{A}_{v}^2\mathcal{A}_{z}^2\right),\nonumber\\
	\mathbb{D}\left\{\omega_2^\mathrm{H}\omega_2\right\}&=\frac{\kappa_G^2}{M^2}\left(2N^2\mathcal{A}_{z}^{(4)}-N^2\mathcal{A}_{z}^2 \right),\nonumber\\
	\mathbb{D}\left\{\omega_3^\mathrm{H}\omega_3\right\}&={\kappa_h^2}N^2.\label{eqn:allcom}
\end{align}
Afterward, substituting Eq. \eqref{eqn:finalw1w2}-\eqref{eqn:allcom} into Eq. \eqref{eqn:varexpand2} yield the final variance expression, while the mean value can be found in Eq. \eqref{eqn:erroronUVZ}. Subsequently, employing a procedure similar to that outlined in Appendix \ref{app:C}, we can derive the shape and rate parameters as illustrated in Lemma \ref{lemma:lemma5}.}

	\bibliography{Reference}
\end{document}